\documentclass[%
 reprint,
superscriptaddress,
nofootinbib,
 amsmath,amssymb,
 aps,
prb,
]{revtex4-1}

\usepackage{graphicx}
\usepackage{dcolumn}
\usepackage{bm}
\usepackage[dvipsnames]{xcolor}
\usepackage[breaklinks=true]{hyperref}
\usepackage[mathlines]{lineno}
\usepackage{setspace} 
\usepackage{miller}
\usepackage{booktabs}
\usepackage{tabularx}
\usepackage{xspace}
\usepackage{physics}

\newcommand{\aFeGe}{\textit{a}-Fe$_x$Ge$_{1-x}$\xspace}
\newcommand{\aFeSi}{\textit{a}-Fe$_x$Si$_{1-x}$\xspace}
\newcommand{\ntot}{$n_{\mathrm{total}}$\xspace}

\renewcommand{\vec}[1]{\mathbf{#1}}

\definecolor{Teal}{RGB}{0, 160, 176}
\definecolor{Red}{RGB}{204, 42, 54}
\definecolor{Orange}{RGB}{235, 104, 65}

\begin{document}


\preprint{APS/123-QED}

\title{Itinerant ferromagnetism and intrinsic anomalous Hall effect \\ in amorphous iron-germanium}

\author{D. S. Bouma}
\affiliation{Department of Physics, University of California, Berkeley, Berkeley, CA 94720, USA}
\affiliation{Materials Sciences Division, Lawrence Berkeley National Laboratory, Berkeley, CA 94720, USA}
\author{Z. Chen}
\affiliation{Materials Sciences Division, Lawrence Berkeley National Laboratory, Berkeley, CA 94720, USA}
\author{B. Zhang}
\affiliation{State Key Laboratory of Surface Physics, Key Laboratory of Computational Physical Sciences,and Department of Physics,
Fudan University, Shanghai 200433, China}
\author{F. Bruni}
\affiliation{Department of Physics, University of California, Berkeley, Berkeley, CA 94720, USA}
\author{M. E. Flatt\'{e}}
\affiliation{Department of Physics and Astronomy, University of Iowa, Iowa City, IA 52242}
\affiliation{Pritzker School of Molecular Engineering, University of Chicago, Chicago, IL 60637}
\affiliation{Department of Applied Physics, Eindhoven University of Technology, P.O. Box 513, 5600 MB Eindhoven, The Netherlands}
\author{A. Ceballos}
\affiliation{Materials Engineering Division, Lawrence Livermore National Laboratory, Livermore, CA 94550, USA}
\author{R. Streubel}
\affiliation{Materials Sciences Division, Lawrence Berkeley National Laboratory, Berkeley, CA 94720, USA}
\author{L.-W. Wang}
\affiliation{Materials Sciences Division, Lawrence Berkeley National Laboratory, Berkeley, CA 94720, USA}
\author{R. Q. Wu}
\affiliation{Department of Physics and Astronomy, University of California, Irvine, Irvine, CA 92627, USA}
\author{F. Hellman}
\affiliation{Department of Physics, University of California, Berkeley, Berkeley, CA 94720, USA}
\affiliation{Materials Sciences Division, Lawrence Berkeley National Laboratory, Berkeley, CA 94720, USA}

\date{\today}

\begin{abstract}
The amorphous iron-germanium system (\aFeGe) lacks long-range structural order and hence lacks a meaningful Brillouin zone. 
The magnetization of \aFeGe is well explained by the Stoner model for Fe concentrations $x$ above the onset of magnetic order around $x=0.4$, indicating that the local order of the amorphous structure preserves the spin-split density of states of the Fe-$3d$ states sufficiently to polarize the electronic structure despite $\vec{k}$ being a bad quantum number.
Measurements reveal an enhanced anomalous Hall resistivity $\rho_{xy}^{\mathrm{AH}}$ relative to crystalline FeGe; this $\rho_{xy}^{\mathrm{AH}}$ is compared to density functional theory calculations of the anomalous Hall conductivity to resolve its underlying mechanisms. 
The intrinsic mechanism, typically understood as the Berry curvature integrated over occupied $\vec{k}$-states but shown here to be equivalent to the density of curvature integrated over occupied energies in aperiodic materials, dominates the anomalous Hall conductivity of \aFeGe ($0.38 \leq x \leq 0.61$). 
The density of curvature is the sum of spin-orbit correlations of local orbital states and can hence be calculated with no reference to $\vec{k}$-space. 
This result and the accompanying Stoner-like model for the intrinsic anomalous Hall conductivity establish a unified understanding of the underlying physics of the anomalous Hall effect in both crystalline and disordered systems.

\end{abstract}

\maketitle

\section{Introduction and Background}
The anomalous Hall effect~\cite{Nagaosa2010} (AHE) refers to a nonzero transverse voltage in zero applied magnetic field. Its underlying physical mechanisms are an enduring topic of research, due both to the insight provided into the effects of electronic structure and impurity scattering on conduction and to the potential spintronic applications of materials hosting a large AHE. 
Specifically, the AHE shares its microscopic origins with the spin Hall effect (SHE), which generates a pure spin current essential for semiconductor spintronics, including the spin-transfer torque manipulation of spin valves, leading to intense experimental interest in materials hosting large AHE as a gateway to large SHE.~\cite{Sinova2015}

The measured Hall resistivity is the sum of two contributions: the ordinary Hall effect (OHE), a potential difference across the sample due to carrier deflection by the Lorentz force, and the anomalous Hall effect (AHE), which is proportional to the magnetization $M_z(H)$ and stems from a combination of asymmetric scattering and band structure effects unique to magnetically ordered materials.\footnote{While the AHE was long understood to be present exclusively in ferromagnets, recent results indicate that certain noncollinear antiferromagnets also host a substantial AHE~[A. K. Nayak \emph{et al.}, Sci. Adv. \textbf{2}, e1501870 (2016)].}
We write this phenomenologically as,~\cite{Pugh1930}
\begin{equation} \label{eqHall}
    \rho_{xy}(H) = \rho_{xy}^{\mathrm{OH}}(H) + \rho_{xy}^{\mathrm{AH}}(H) = R_0 H + R_S M_z(H),
\end{equation}
where $M_z(H)$ is the $z$-component of the magnetization measured with the field $H$ perpendicular to the plane of the sample, and $R_0$ and $R_S$ are the ordinary and anomalous Hall coefficients, respectively. $\rho_{xy}^{\mathrm{OH}}$ contributes a linear in $H$ ``background'' whose slope shares the sign of the carriers in the sample, while $\rho_{xy}^{\mathrm{AH}}$ adopts the shape of $M_z(H)$. 

As understanding of the AHE deepened, analytical focus shifted from the anomalous Hall resistivity $\rho_{xy}^{\mathrm{AH}}$ to the more fundamental anomalous Hall conductivity (AHC) $\sigma_{xy}^{\mathrm{AH}}$, which when $\rho_{xy}^{\mathrm{AH}}$ is small relative to the longitudinal conductivity $\rho_{xx}$ is given by $\sigma_{xy}^{\mathrm{AH}} \approx \rho_{xy}^{\mathrm{AH}}/\rho_{xx}^2$. The AHC originates from a combination of three mechanisms: skew-scattering, side-jump scattering, and the intrinsic mechanism. 
Skew-scattering refers to asymmetric scattering of spin-polarized carriers by spin-orbit-coupled impurities;~\cite{Smit1955} while side-jump scattering refers to the transverse velocity deflection caused by the opposite electric fields experienced by spin-polarized carriers upon approaching and leaving a spin-orbit-coupled impurity.~\cite{Berger1970} 
The intrinsic contribution is proportional to the integral over occupied states of the Berry curvature $\Omega(\vec{k})$, which can be written as a summation over eigenstates:~\cite{Xiao2010} 
\begin{equation} \label{eq:BerryCurvature}
    \Omega(\vec{k}) = i \sum_{n \neq n'}^{} \frac{\bra{n}\grad_{\vec{k}}\hat{H}\dyad{n'}\grad_{\vec{k}}\hat{H}\ket{n}} {(\varepsilon_n - \varepsilon_{n'})^2}.
\end{equation}
Eq.~\ref{eq:BerryCurvature} emphasizes that this intrinsic contribution to the AHC depends only on the topology of the band structure and not on the details of scattering in the material.~\cite{Jungwirth2002, Onoda2002} 

The relative importance of these three mechanisms has long been disputed, but some clarity has come from a unified theory, originally proposed by Onoda~\cite{Onoda2006} and elaborated upon by Liu,~\cite{Liu2011} which separates systems into three regimes based on longitudinal conductivity $\sigma_{xx} = \rho_{xx}^{-1}$: a high conductivity ``clean'' limit where $\sigma_{xx} \gtrsim 10^6 \,(\mathrm{\Omega cm})^{-1}$ and $\sigma_{xy}^{\mathrm{AH}} \propto \sigma_{xx}$, indicating that the AHC is dominated by skew-scattering; an intermediate ``good metal'' regime where $10^4 \lesssim \sigma_{xx} \lesssim 10^6 \,(\mathrm{\Omega\; cm})^{-1}$ and $\sigma_{xy}^{\mathrm{AH}}$ is independent of $\sigma_{xx}$, indicating that the intrinsic mechanism dominates the AHC; and a low conductivity regime where $\sigma_{xx} \lesssim 10^4 \,(\mathrm{\Omega\; cm})^{-1}$, which was less well understood. 
At the time of Onoda's work, this last regime was described empirically as having $\sigma_{xy}^{\mathrm{AH}} \sim \sigma_{xx}^{\gamma}$, with $\gamma$ ranging between $1.6-1.8$; Onoda recovered the scaling $\gamma = 1.6$ in the hopping regime, and Liu found $1.33 \leq \gamma \leq 1.76$ by studying multiple hopping processes which typically occur deep in the low conductivity regime ($\sigma_{xx} \lesssim 10^2 \,(\mathrm{\Omega\; cm})^{-1}$), demonstrating that the microscopic origin of the AHE determines its scaling in this insulating regime. 

Recent experimental results from Karel \textit{et al.}~\cite{Karel2016} explored the high conductivity edge of the low conductivity regime, employing an empirical scaling argument based on the intrinsic AHC in \textit{crystalline} Fe to suggest that the AHC in the low conductivity metal amorphous iron-silicon (\aFeSi) is dominated by the intrinsic mechanism. 
This conclusion is intially surprising, because perturbative expressions such as Eq.~\ref{eq:BerryCurvature} indicate that the intrinsic anomalous Hall effect is maximized when the spin-orbit interaction splits band dispersions to create anti-crossing points with a small gap near the Fermi level, which resonantly enhances the integral in question~\cite{Onoda2006} by enhancing the momentum-space Berry curvature at these points. 
However, full-band calculations on crystalline materials have called into question assumptions based on perturbative treatments of anti-crossings near the Fermi level;~\cite{Sahin2015} in addition, for an amorphous system, $\vec{k}$ is no longer a good quantum number due to the absence of lattice periodicity. We will instead introduce the energy-resolved density of Berry curvature~\cite{Sahin2015, Guo2008} for an amorphous system and show that it supplants a typical Brillouin-zone integral to allow a more general expression of the electronic origins of the intrinsic AHC.
In a quite distinct but relevant approach, Marrazzo and Resta~\cite{Marrazzo2017a} have reformulated the expression for the intrinsic AHC in terms of real-space wavefunctions, enabling the calculation of this contribution in systems where $\vec{k}$ is not a good quantum number, including structurally disordered systems such as \aFeSi. 

Here we study amorphous iron-germanium (\aFeGe), which promises similarly tunable magnetization and resistivity to \aFeSi, but accompanied by spin-orbit coupling (SOC) large enough to fundamentally alter both the electronic structure and the magnetic texture in \aFeGe by amplifying the antisymmetric Dzyaloshinskii-Moriya interaction (DMI).
While recent investigation in the Fe-Ge system has focused on the  magnetic skyrmion lattice in single crystals and epitaxial films of cubic B20 FeGe,~\cite{Huang2012, Porter2014, Gallagher2016a} even prior to the skyrmion discovery, the Fe-Ge system had established a record of varied spin configurations that depend intimately on structure and composition: hexagonal B35 FeGe hosts a conical structure at low temperature, which gives way to a collinear antiferromagnet at higher temperatures; monoclinic FeGe is also a collinear antiferromagnet; tetragonal FeGe$_2$ hosts a collinear antiferromagnet at low temperatures which transitions to a noncollinear antiferromagnet at higher temperatures; finally, both hexagonal and cubic Fe$_3$Ge are ferromagnetic. 
Amorphous Fe-Ge films have previously been fabricated;~\cite{Suran1976, Daver1977, Terzieff1979, Randhawa1981, Lorentz1984} these films showed ferromagnetism above a critical Fe concentration $x \approx 0.4$, and in one case preliminary indications of noncollinear spin textures were observed in Lorentz transmission electron micrographs.~\cite{Randhawa1981}
However, as much of this work predates the discovery of the Berry phase and the ensuing resurgence of interest in the anomalous Hall effect, magnetotransport in this material has yet to be thoroughly explored.

In this work we conduct experimental and computational investigations of the low temperature ($T=2$ K) magnetization and magnetotransport including Hall resistivity in \aFeGe  with $0.38 \leq x \leq 0.61$, contextualizing our findings via the isoelectronic \aFeSi system to highlight broader trends among amorphous ferromagnets. 
Ultimately, we show that \aFeGe shares considerable physics with \aFeSi and, unlike \aFeSi, with its crystalline counterparts. 
We supplement the empirical scaling argument with a new theoretical paradigm for the intrinsic anomalous Hall conductivity in aperiodic materials, based on the energy-resolved density of Berry curvature, and demonstrate how this model and our data together corroborate the intrinsic origin of the anomalous Hall effect in the low conductivity regime. 
Finally, we comment on the implications of our results for future work on other spin- and orbitronic phenomena arising from the Berry curvature.

\section{Methods}

\begin{figure*}
\centering
\includegraphics[height = 1.75in]{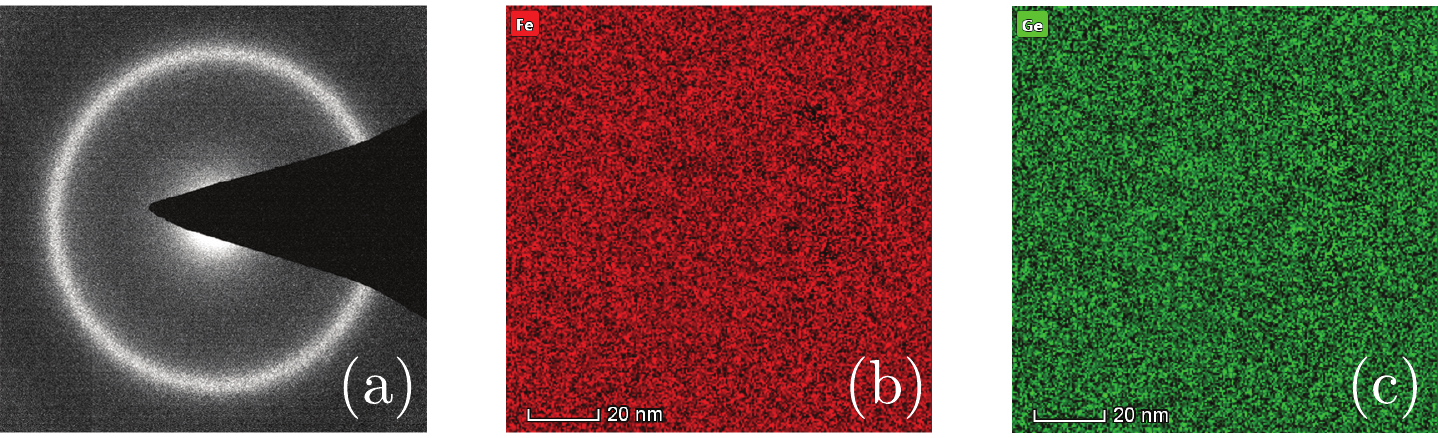}
\caption{Structural characterization of \aFeGe thin films.  Transmission electron diffractogram (a) taken with a 200nm probe on $x = 0.61$ shows the diffuse ring characteristic of an amorphous material. High resolution energy dispersive X-ray spectrometry images with 20 nm scale bar show nanoscale elemental homogeneity in the distribution of (b) Fe and (c) Ge for $x = 0.56$. The pixellation in these images is the sub-nm data resolution limit.}
\label{fStruct}
\end{figure*}

\subsection{Thin Film Growth and Characterization}
Films of amorphous iron-germanium (65-80 nm  thickness) were deposited onto commercially available amorphous silicon nitride substrates ($5000 \mathrm{\AA}$ LPCVD Si$_3$N$_4$/$300 \mathrm{\AA}$ SiO$_2$/Si $\hkl<1 0 0>$ wafer) by co-evaporation of iron from an electron beam source and germanium from an effusion cell in a chamber with base pressure below $8 \times 10^{-9}$ Torr.  
Growth rates were monitored using separate quartz crystal microbalances and were, for each deposition, maintained at a constant value  between $0.23$ \AA/s and $0.27$ \AA/s. Immediately following film growth, a capping layer (3-5 nm thickness) which prevents sample oxidation was deposited \emph{in situ} by sublimation of Al$_2$O$_3$ from an electron beam source.

Film thickness was measured in a KLA-Tencor Alpha-Step IQ surface profiler. Compositions and atomic number densities were determined from Rutherford backscattering spectra measured at the Pelletron at Lawrence Berkeley National Laboratory; a spectrum corresponding to each measurement was simulated and iterated to fit the measured spectrum to obtain the sample composition and density. 
Sample amorphicity was verified using x-ray and electron diffraction [Fig.~\ref{fStruct}(a)], while high-resolution energy dispersive X-ray spectrometry [Fig.~\ref{fStruct}(b) and (c)] demonstrates the films' nanoscale elemental homogeneity. 
Taken together, these data rule out the possibility of precipitation, percolation, or nanocrystallization of the \aFeGe films.
Furthermore, the amorphous quality of \aFeSi films grown by identical techniques has been previously shown using high-resolution cross-sectional transmission electron microscopy.~\cite{Gray2011} 
Given that silicon and germanium are both well-known glass formers, it is unsurprising that both \aFeSi and \aFeGe readily condense into an amorphous phase in the studied range of $x$.

Magnetization and magnetoelectrical transport were measured in a Quantum Design Magnetic Property Measurement System (MPMS XL) equipped with a 7 T superconducting magnet.  The reciprocating sample option (RSO) was used for all magnetization measurements, with four distinct SQUID voltage curves averaged for each magnetic moment data point. To isolate the magnetic signal from the film, a diamagnetic background due to the substrate was calculated from the sample mass, measured in a microbalance, and substrate susceptibility, known from previous magnetization measurements of virgin substrates. This calculated background was used to verify the accuracy of a linear fit to the high-field regions of a $T = 300$ K $M(H)$ measurement on each sample, and a straight line with the slope of this fit was subtracted from the total measured magnetization. 

Longitudinal and transverse resistivities were obtained using the van der Pauw method,~\cite{VanderPauw1958} in which each rectangular film (side length $3-5$ mm) is mounted onto a specialized MPMS sample rod, with four evenly spaced pointlike (diameter $< 0.5$ mm) electrical contacts attached to the film perimeter by indium soldering. Individual four-point resistances used to calculate the film resistivity were measured using standard AC lock-in techniques, with currents of amplitude $2\; \mu$A and frequency 16 Hz. The transverse and longitudinal resistivities were then calculated following the procedures detailed by van der Pauw, with the aid of a numerical solver to find roots of the transcendental equation defining the longitudinal resistivity.

\subsection{Density Functional Theory Calculations}
Density functional theory (DFT) calculations were performed using the projector augmented wave (PAW) method~\cite{Blochl1994, Kresse1999} and a plane wave basis set, as implemented in the Vienna \textit{ab initio} simulation package (VASP);~\cite{Kresse1993, Kresse1996} an energy cutoff of 350 eV was used for the plane wave basis expansion.
The exchange-correlation interactions were described by the generalized-gradient approximation (GGA) with the Perdew-Burke-Ernzerhof (PBE) functional,~\cite{Perdew1996} taking spin polarization into account.
Two separate sets of amorphous Fe$_x$Ge$_{1-x}$ structures were simulated, with one set using a 64-atom supercell (with 29, 31, 35, and 39 Fe atoms) and the other using a 128-atom supercell (with 58, 61, 69, and 78 Fe atoms), corresponding to $x \simeq 0.45$, 0.48, 0.54, and 0.61, matching the experimental compositions. The similarity of the results of these calculations indicates that both supercell sizes capture the patterns of amorphous systems. 

For each value of $x$ and each supercell size, 20 different independent amorphous structures were simulated using \textit{ab initio} molecular dynamics (AIMD).
Initial configurations were created by randomly substituting Fe and Ge atoms onto zincblende crystal structure sites. 
To randomize the atomic structural positions, the systems underwent a melting step (2000 K for the 64-atom systems; 3000 K followed by a 4 ps anneal to ensure a fully molten state for the 128-atom systems) and a quenching step (2000 K to 200 K at $3 \times 10^{14}$ K/s for the 64-atom systems; 3000 K to 300 K at $4.5 \times 10^{14}$ K/s for the 128-atom systems); the 64-atom systems underwent a subsequent annealing step (200 K for 4.5 ps) in a canonical ensemble. 
The structures were then further relaxed using the conjugate gradient method until the forces on each atom were less than 0.01 eV/\AA. Only the $\Gamma$-point was used to sample the Brillouin zone during the melting, quenching, and annealing processes; however, a $3 \times 3 \times 3$ Monkhorst-Pack (MP) $k$-point mesh was used for the atomic geometry relaxation of all structures. 
This mesh was also used for the 128-atom electronic structure calculation, while the 64-atom electronic structure calculation employed a $6 \times 6 \times 6$ MP $k$-point mesh.

\begin{figure}
    \includegraphics{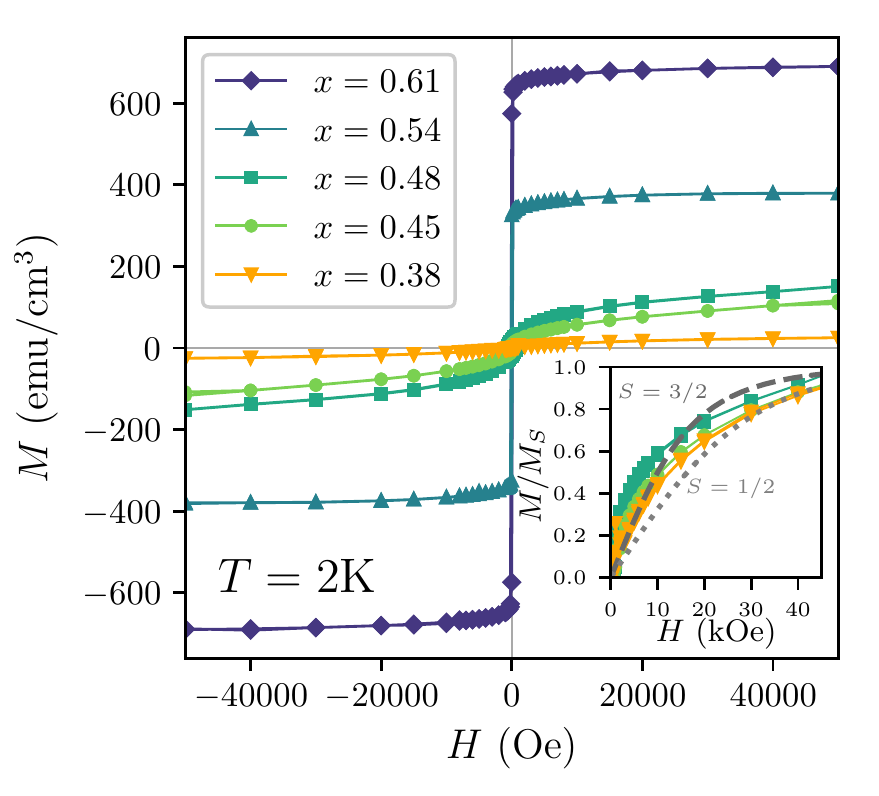}
    \caption{$M(H)$ curves for films with $0.38 \leq x \leq 0.61$ measured at $T = 2$ K with $H$ parallel to the film plane.  A temperature independent diamagnetic background corresponding to the combined magnetization of each sample's substrate and capping layer as measured at $T=300$ K has been subtracted from each measurement. Inset: expanded view of the $M(H)$ curves with $x < 0.50$, each normalized to its saturation magnetization as defined in the text and overlaid with Brillouin functions corresponding to $S = 3/2$ and $S = 1/2$. Data points are connected as a guide to the eye.}
    \label{f1}
\end{figure}

\section{Results}
\subsection{Magnetization}

Experimental magnetization curves at $T = 2$ K with the applied field $H$ in the plane of the film are shown in Fig.~\ref{f1} for \aFeGe with $x=0.38$, 0.45, 0.48, 0.54, and 0.61. 
While the M(H) curves for $x > 0.50$ primarily exhibit a squareness and absence of hysteresis associated with an otherwise unremarkable soft ferromagnet, we note the curvature at $H < 2500$~Oe. For lower $x$ ($x = 0.48$ and 0.45) this curvature is dramatically exaggerated and the curve takes on an S-shape while also developing a small hysteresis at low fields. 
The low-$H$ curvature at high $x$ and S-shaped curves at lower $x$ hint at non-collinear spin textures in \aFeGe ($0.45 \leq x \leq 0.61$). Such spin textures have since been observed directly in this system via resonant X-ray scattering and Lorentz transmission electron microscopy; detailed results will be reported elsewhere.~\cite{Chen2019,Streubel2019}



\begin{figure}
    \includegraphics{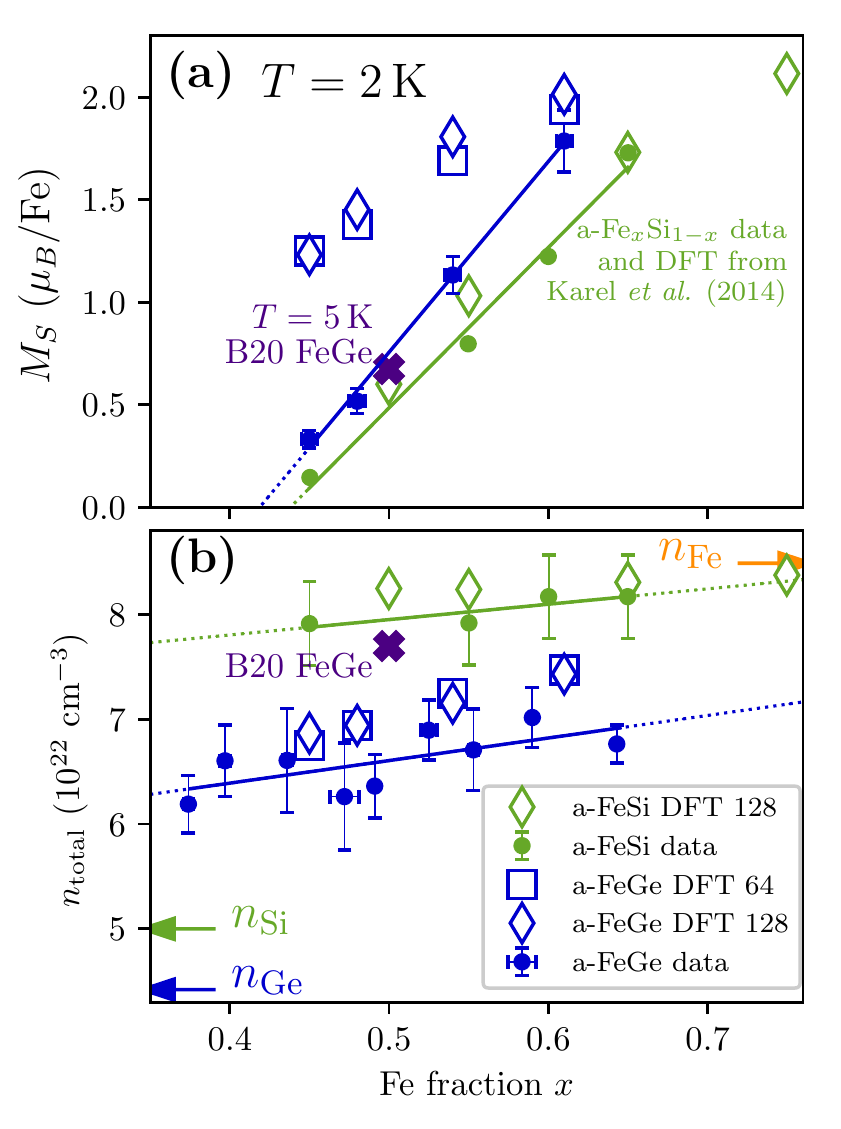}
    \caption{Comparison of magnetic and structural properties of \aFeGe (blue symbols) and \aFeSi (green symbols) indicates that both materials exhibit a Stoner model-type itinerant ferromagnetism. The purple $\mathbf{\times}$ marks $T = 5$ K measurements of B20 FeGe for comparison. (a) Magnetization per Fe atom (filled circles) increases roughly linearly as a function of Fe concentration $x$ for both \aFeGe and \aFeSi. MD-DFT calculations of \aFeGe using a 64-atom supercell (open blue squares) yield the same values of $M_S$ as those using a 128-atom supercell (open blue diamonds); these are discussed further in the text. (b) Total atomic number density increases slightly with $x$; solid lines are linear fits to experimental data and dotted lines are extrapolations of those fits shown as guides to the eye. \aFeSi MD-DFT calculations and $T = 2$ K experimental data are from Ref.~\citenum{Karel2014a}, and $T = 5$ K B20 FeGe experimental data is from Ref.~\citenum{Gallagher2016a}.}
    \label{f2}
\end{figure}

Although the $M(H)$ loops for \aFeGe are qualitatively similar to \aFeSi~\cite{Karel2016} for the same $x$, the volume magnetization is slightly greater in \aFeGe and the S-shape of the magnetization in the low-$x$ samples is exaggerated, indicating a more complex spin structure in \aFeGe that persists to higher fields.
The measured and calculated values of the $T=2$~K saturation magnetization per Fe atom of \aFeSi and \aFeGe are compared quantitatively in Fig.~\ref{f2}(a), where the data and calculations for \aFeSi are from Ref.~\citenum{Karel2014a}. 
The \aFeGe saturation magnetization was obtained for high $x$ (reasonably square loops in Fig.~\ref{f1}) by extrapolating the $M(H)$ curve for $H \geq 4$ T to $H = 0$; this extrapolation yields almost exactly $M_S = M(H = 5\mathrm{\,T})$ so the $H = 5$ T value of $M$ was chosen as $M_S$ for the low $x$ measurements (S-shaped loops in Fig.~\ref{f1}). 
The enhanced $M_S$ in \aFeGe compared to \aFeSi is supported by the presence of sharper peaks in the calculated spin-resolved density of states in \aFeGe, which will be further discussed later in the paper.

\begin{figure*}
    \includegraphics{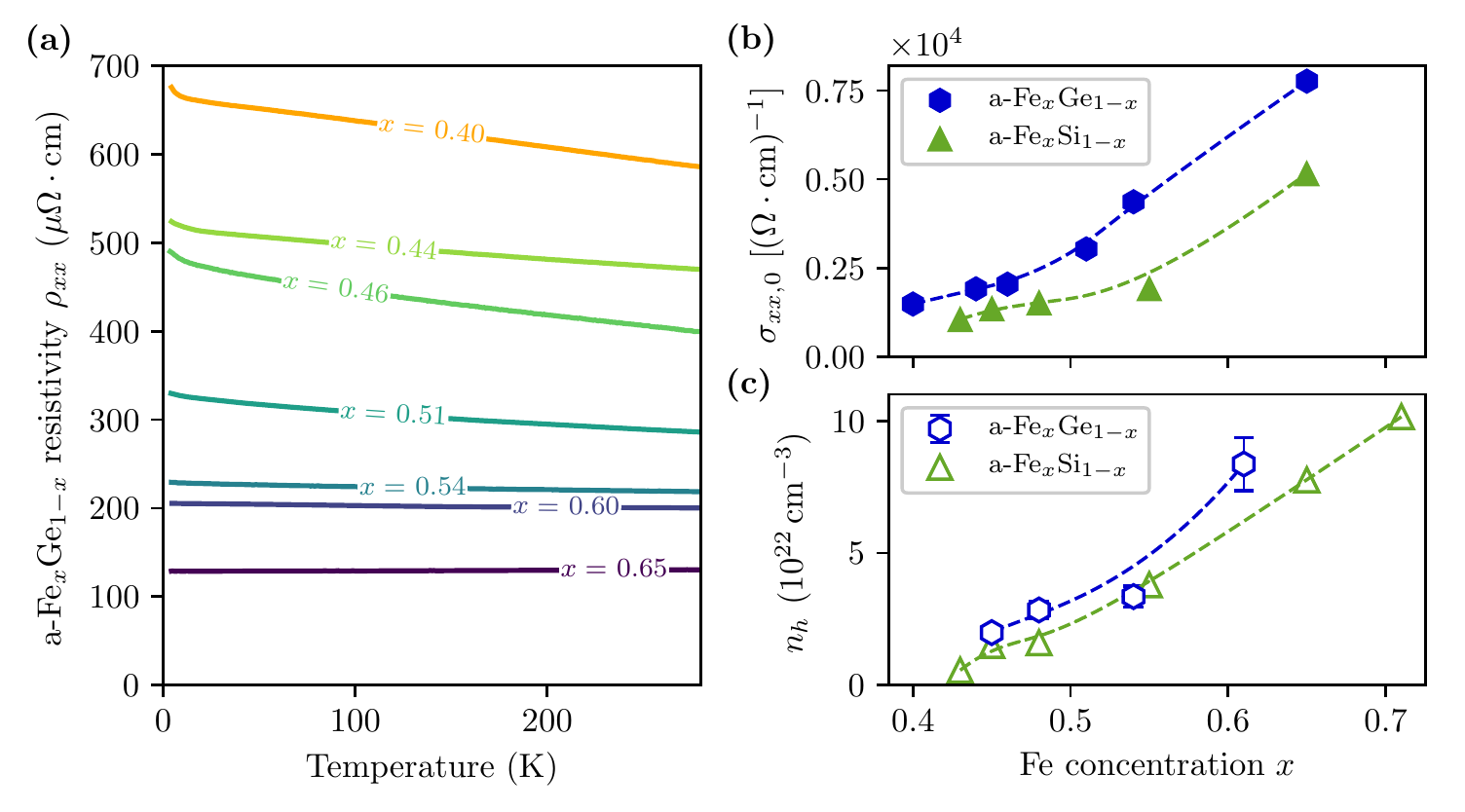}
    \caption{Longitudinal charge transport properties of \aFeGe (blue lines and hexagons), shown with \aFeSi data from Ref.~\citenum{Karel2016} (green triangles) for comparison. (a) Longitudinal resistivity $\rho_{xx}$ as a function of temperature $T$ exhibits a negative temperature coefficient of resistance for all compositions $x$, and decreases with increasing $x$ at all $T$. Individual data points are too closely spaced to be distinguishable, so a solid line is shown instead. (b) Residual longitudinal conductivity $\sigma_{xx,0}$ measured at $T = 4 \,\mathrm{K}$ as a function of Fe concentration $x$ for \aFeGe (solid blue hexagons) and \aFeSi (solid green triangles). Spline fits to the data (dashed lines) are shown as a guide to the eye. (c) Carrier concentration $n_h$ obtained from Hall effect measurements at $T = 2$ K for \aFeGe (open blue hexagons) and \aFeSi (open green triangles). Spline fits to the data (dashed lines) are shown as a guide to the eye. }
    \label{fRho}
\end{figure*}

\begin{figure}
    \includegraphics{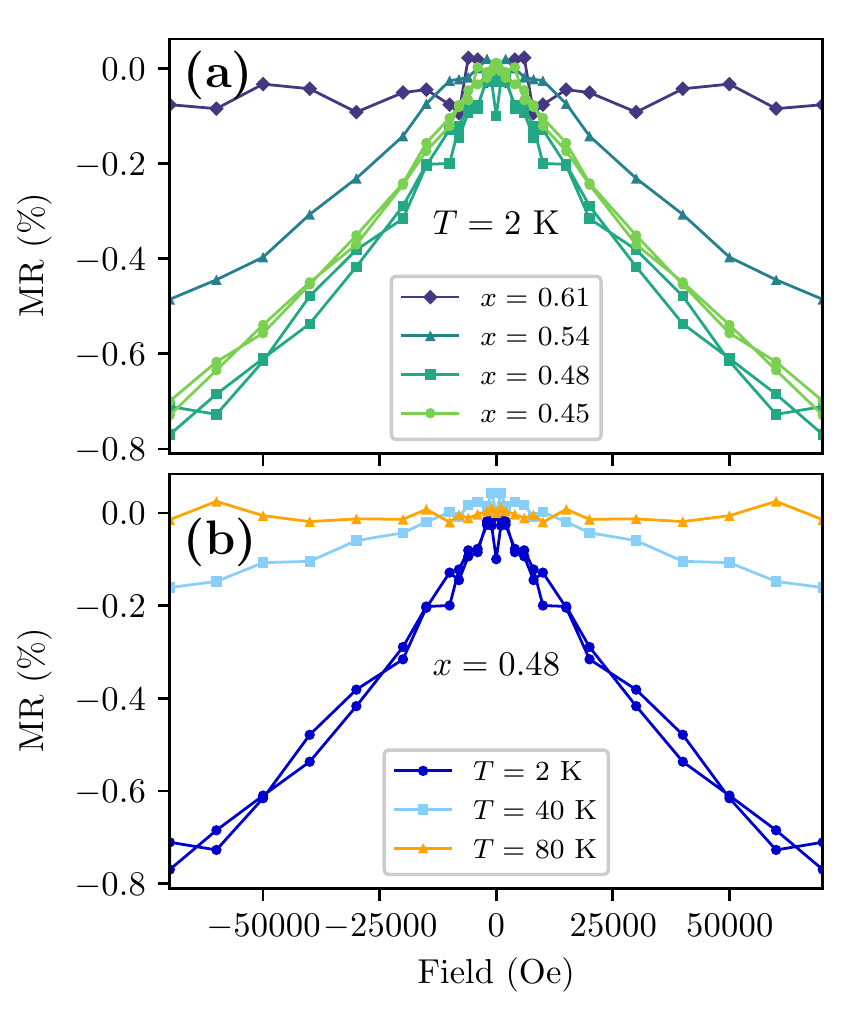}
    \caption{Normalized magnetoresistance with $H$ oriented out of the film plane for \aFeGe. Data is symmetrized according to $\rho(H) = \frac{1}{2}[\rho(+H) + \rho(-H)]$ to remove contact asymmetry effects; data points are connected as a guide to the eye. Poor signal-to-noise is a side effect of the measurement geometry, which was chosen to optimize the Hall voltage signal. (a) Composition dependence of magnetoresistance at $T = 2$ K shows stronger negative magnetoresistance for lower $x$. (b) Temperature dependence of magnetoresistance for $x = 0.48$ shows negative magnetoresistance to $T = 40$ K, disappearing by $T = 80$ K.}
    \label{fMR}
\end{figure}

Fig.~\ref{f2}(b) shows measured and calculated values of the atomic number density \ntot for \aFeGe and \aFeSi; \ntot for crystalline B20 FeGe is also shown for comparison. 
The increased size of the Ge atom compared to the Si atom is responsible for the 15-20\% reduction in \ntot of \aFeGe compared to that of \aFeSi.
Since Fe structures itself more compactly than either Si or Ge, \ntot unsurprisingly increases with increasing $x$ for both systems; DFT calculations reproduce this trend in good qualitative agreement with experiment for both \aFeSi and \aFeGe. 
Cursory extrapolations of the measured trend indicate that, for this $x$ regime, \ntot for both systems approaches the atomic number density of bulk Fe ($n_{\mathrm{Fe}}$, orange arrow on right vertical axis) as $x$ approaches 1. 
However, neither extrapolation recovers its group IV element atomic number density (arrows on left vertical axis: green for $n_{\mathrm{Si}}$ and blue for $n_{\mathrm{Ge}}$) as $x$ approaches 0, suggesting the existence of one or more polyamorphous transitions in both systems.




\subsection{Magnetotransport}
\begin{figure}
    \includegraphics{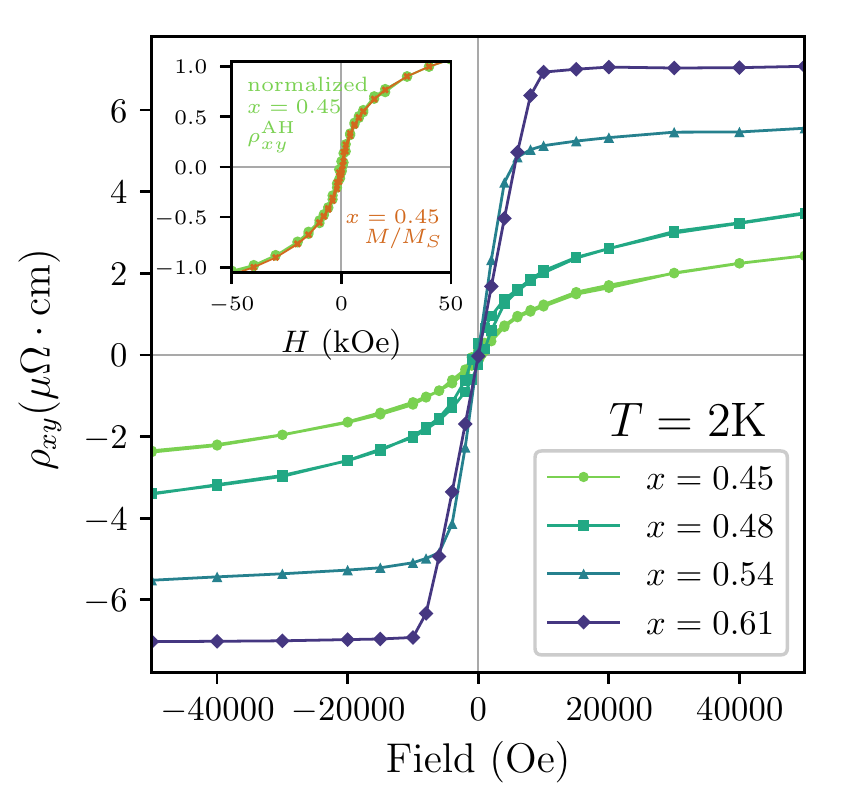}
    \caption{Measured Hall resistivity $\rho_{xy}$ as a function of applied magnetic field at $T = 2 \mathrm{K}$ for various Fe concentrations $x$ of \aFeGe shows a dominant anomalous Hall effect with a very large magnitude for all $x$. Inset: normalized anomalous Hall resistivity $\rho_{xy}^{\mathrm{AH}} (H) \equiv \rho_{xy} (H) - R_0 H$ (green) overlaid with normalized measured magnetization (orange), showing excellent agreement and further corroborating the weak ferromagnetism in the $x=0.45$ and $0.48$ films. Data points are connected as a guide to the eye.}
    \label{f3}
\end{figure}

Longitudinal charge transport measurements are shown in Figs.~\ref{fRho} and~\ref{fMR}, while transverse measurements are shown in Fig.~\ref{f3}; we first examine the $H = 0$ longitudinal transport properties. 
Fig.~\ref{fRho}(a) shows the temperature dependence of the longitudinal resistivity $\rho_{xx}$, or simply $\rho$, for \aFeGe with $0.40 \leq x \leq 0.65$, with higher $x$ values in darker colors (solid lines are shown because the individual data points are too closely spaced to be distinguishable). 
The temperature dependence of $\rho$ for each $x$ is relatively weak compared to the composition dependence of $\rho$ across the range of $x$ studied. 
Both the weak temperature dependence of the resistivity and the negative sign of the temperature coefficient of resistance $\alpha$ ($\alpha \equiv \rho^{-1} \, \partial \rho / \partial T$) are consistent with the temperature-independent scattering length $\ell \sim a$ of an amorphous metal (where $a$ is the average interatomic spacing); this temperature-independent $\ell$ leads to the condition $\rho \gtrsim 150 \, \mu\Omega\cdot\mathrm{cm}$ for amorphous metals, which is apparent in our measurements. 
The systematic increase in $\rho$ with decreasing $x$ reflects the decrease in carrier concentration with decreasing $x$, which will be further discussed next.

Fig.~\ref{fRho}(b) shows the residual conductivity $\sigma_{xx,0}(T = 4\,\mathrm{K})$ while Fig.~\ref{fRho}(c) shows the carrier concentration for \aFeGe compared to \aFeSi (\aFeSi data from Ref.~\citenum{Karel2016}).
The carrier concentration was obtained from magnetotransport measurements carried out at $T = 2$ K (Fig.~\ref{f3}) with the samples set up in the van der Pauw geometry. 
The anomalous Hall effect dominates $\rho_{xy}$ for all four samples; the additional positive slope due to the ordinary Hall effect (indicative of hole carriers) is apparent for the $x=0.45$ and $0.48$ samples, but is hardly visible at higher $x$ so the density of holes $n_h$ was extracted from $\rho_{xy}(H)$ by iteratively fitting the parameters $R_0$ and $R_S$ to Eq.~\ref{eqHall} in Python, using measured $M_z(H)$ data for each sample. 
$R_0$ is related to the carrier concentration by $|R_0| = (ne)^{-1}$, where $n$ is the carrier concentration and $e$ is the electronic charge; our fits yield the values of hole concentration $n_h$ shown by open blue hexagons in Fig.~\ref{fRho}(c). 
Comparing our results to the measured (for $0.43 \leq x \leq 0.48)$ and extrapolated (for $x > 0.48$) values of $n_h$ for \aFeSi (open green triangles) confirms that the higher carrier concentration in \aFeGe is responsible for its higher $\sigma_{xx,0}$ for any $x$, represented by filled blue hexagons in Fig.~\ref{fRho}(b), compared to \aFeSi (filled green triangles).

The magnetoresistance (MR) ratio, defined as $\mathrm{MR}(H) \equiv [\rho(H) - \rho(0)]/\rho(0)$, was also measured for \aFeGe and is shown in Fig.~\ref{fMR}(a) at $T = 2$ K with $H$ oriented perpendicular to the film plane. 
As expected for ferromagnets, the MR is negative,~\cite{Spaldin2010} and the weakly ferromagnetic $x = 0.45$ and 0.48 samples show a larger negative magnetoresistance due to increased suppression of spin-disorder scattering by an applied magnetic field. 
Fig.~\ref{fMR}(b) shows that, in contrast to the positive MR of \aFeSi for $T \geq 16$ K,~\cite{Karel2016} this significantly enhanced negative contribution to the MR of \aFeGe dominates up to temperatures between 40 K and 80 K, resulting in a net negative MR for $x = 0.48$ over a wide temperature range.
We note that, while still small, the magnitude of the $T = 2$ K MR in \aFeGe is more than 200 times greater than the effect measured in the same geometry (H perpendicular to the film and therefore also perpendicular to the current) in \aFeSi for comparable Fe concentrations, despite similar $M(H)$. 
Since we have shown our samples to be structurally and chemically homogeneous (Fig.~\ref{fStruct}), we attribute this enhanced negative MR in \aFeGe both to a stronger coupling between carriers and local moments and to a greater reduction in the disorder seen by the carriers. 
The latter effect reinforces our observation, based on the $M(H)$ curves, that \aFeGe hosts a more complicated spin texture than \aFeSi.

We isolate the anomalous component of the Hall resistivity by subtracting the OHE from the total $\rho_{xy}$, and show the result for $x=0.45$ in the inset to Fig.~\ref{f3} normalized by its $H=5$ T value and overlaid with the correspondingly normalized $M_z(H)$ to emphasize the unmistakable magnetic origin of this Hall signal, even in our most weakly ferromagnetic sample. 
This inset also exemplifies the absence of additional measurable contributions to the Hall effect, such as a topological Hall effect that has been observed in several systems as a consequence of the adiabatic change in the carrier phase by a noncollinear spin texture. 
While the samples shown here likely host local noncollinear spin textures due to the Dzyaloshinskii-Moriya interaction in the Fe-Ge system, the absence of global chirality in the amorphous structure reduces any net signal from localized topological contributions to the Hall resistivity below the sensitivity of our setup.

\begin{figure}
    \includegraphics{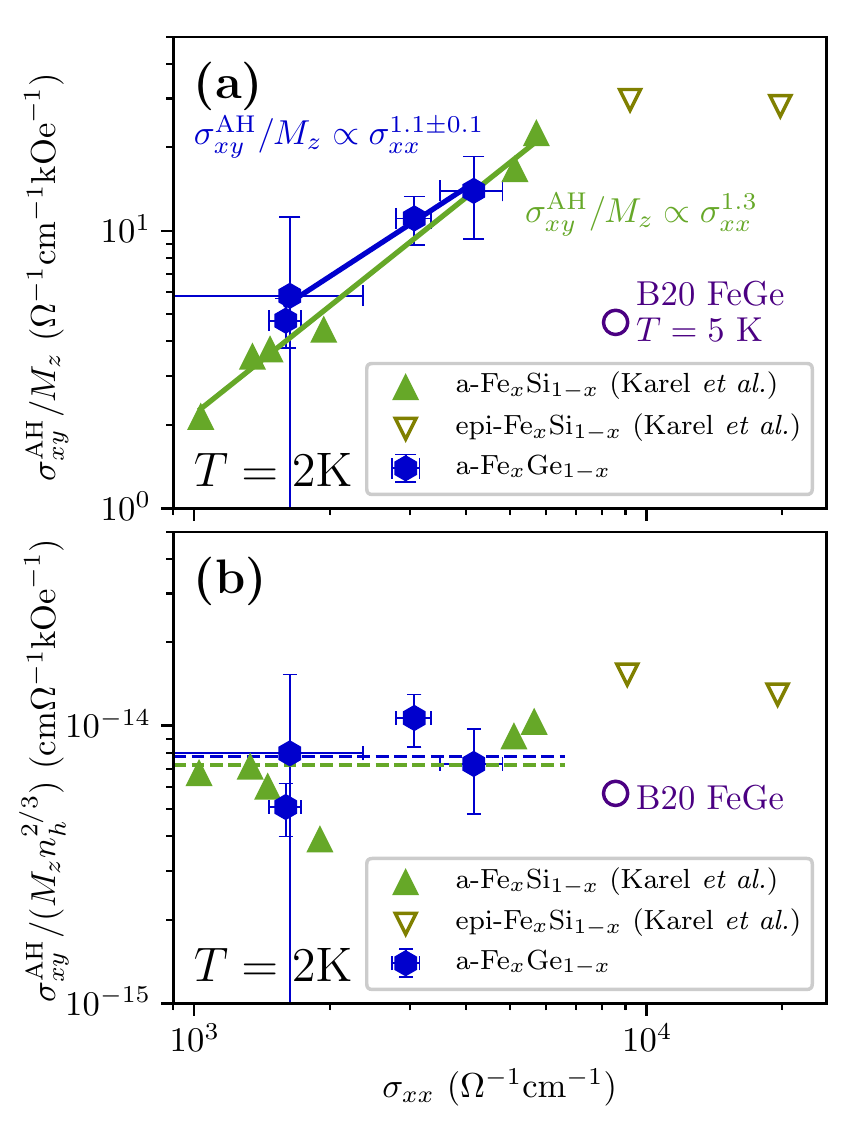}
    \caption{Scaling plot of the anomalous Hall conductivity $\sigma_{xy}^{\mathrm{AH}}$ reveals the origin of the anomalous Hall effect in \aFeSi and \aFeGe across different regimes of longitudinal conductivity $\sigma_{xx}$. (a) Normalizing $\sigma_{xy}^{\mathrm{AH}}$ by the saturation magnetization $M_z$ yields a power law dependence that deviates from the empirical $\sigma_{xy}^{\mathrm{AH}} \sim \sigma_{xx}^{1.6-1.8}$. (b) Normalizing further by a free-electron-type dependence on the carrier concentration, $n_h^{2/3}$, yields an essentially constant value for both \aFeSi and \aFeGe. Dashed lines are average values of $\sigma_{xy}^{\mathrm{AH}}/(M_z n_h^{2/3})$ for \aFeGe (blue) and \aFeSi (green). Amorphous and epitaxial Fe-Si data from Ref.~\citenum{Karel2016}; B20 FeGe data from Ref.~\citenum{Gallagher2016a}.}
    \label{f4}
\end{figure}



\section{Discussion}
\subsection{Magnetization}
The onset of magnetic order at $T = 2$ K in \aFeGe occurs between $x = 0.45$ and $x = 0.38$, in agreement with previous work.~\cite{Suran1976} The inset of Fig.~\ref{f1} confirms this by comparing the magnetization of the $x = 0.38,\,0.45,$ and $0.48$ films to Brillouin functions of ideal paramagnets with $S = 1/2$ and $S = 3/2$; the paramagnetic $x=0.38$ curve falls squarely between the Brillouin functions while no single Brillouin function adequately models both the low- and high-field behavior of the $x=0.45$ and $x=0.48$ curves. The steep slope and finite hysteresis of their magnetization at low fields suggests ferromagnetism, but the magnetization continues to increase without saturating to fields above $H = 7$ T, leading to the conclusion that \textit{a}-Fe$_{0.45}$Ge$_{0.55}$ and \textit{a}-Fe$_{0.48}$Ge$_{0.52}$ are only weakly ferromagnetic.
Notably, the Ising model from which the Brillouin functions arise does not capture the profusion of spin textures that form across the crystalline Fe-Ge system thanks to the competition between Heisenberg and DM interactions in different lattice configurations, and the non-square features of the hysteresis loops noted previously indicate that this competition persists in \aFeGe.

For both \aFeGe and \aFeSi, the magnetization per Fe atom increases with increasing Fe concentration $x$; however, for any given $x$, \aFeGe has a significantly higher  magnetization per Fe atom than \aFeSi, despite having only slightly higher magnetization per unit volume, due to the different atomic densities of these two alloys at all $x$ [Fig.~\ref{f2}(b)]. 
In contrast to the Fe-Si system, in which the magnetization of the crystalline phase is an order of magnitude (or more) less than than that of \aFeSi,~\cite{Karel2014a} the measured magnetization per Fe atom of B20 FeGe at $T = 5$ K from Ref.~\citenum{Gallagher2016a} aligns very well with our experimental data for \aFeGe; taken together with existing evidence that the local atomic environment strongly influences the magnetic state of amorphous transition metal germanides and silicides, this suggests that the local atomic environment in our amorphous films resembles that of B20 FeGe.
The DFT calculations of magnetization in \aFeSi~\cite{Karel2014a} and \aFeGe [open diamonds and squares in Fig.~\ref{f2}(a)] reproduce this experimental trend as well as the relative values of $M$ between both systems, but yield excellent quantitative agreement with experiment only in \aFeSi. 
The discrepancy between calculated and measured magnetization in \aFeGe is likely due to a noncollinear spin texture, which suppresses the net magnetization and is not computationally taken into account because of the difficulty of calculating the spin texture of an amorphous system using DFT. 

\begin{figure}
    \includegraphics{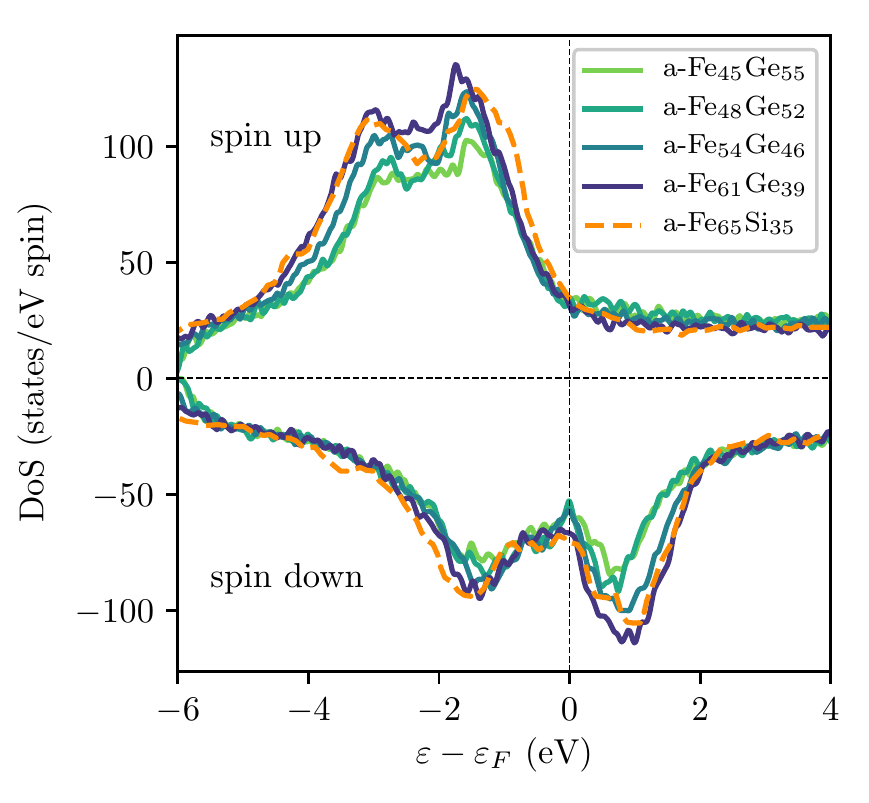}
    \caption{Calculated density of states (DOS) near the Fermi level for the compositions $x$ of \aFeGe studied experimentally (blue curves), and for \textit{a}-Fe$_{0.65}$Si$_{0.35}$ (green dashed curve, from Ref.~\citenum{Karel2014a}), all based on a 128-atom unit cell. The peaks of the majority spin in the DOS are narrowed and sharpened by substituting the larger Ge atom for the Si atom for a fixed $x$, as can be seen from comparing the DOS of \textit{a}-Fe$_{0.65}$Si$_{0.35}$ to that of \textit{a}-Fe$_{0.61}$Ge$_{0.39}$, explaining the enhanced magnetization in \aFeGe.}
    \label{fDoS}
\end{figure}

The increased size of the Ge atom relative to the Si atom enhances the magnetization of \aFeGe relative to \aFeSi [Fig.~\ref{f2}(a)].
A local-moment picture does not fully capture the physics in these systems, so we turn to the calculated densities of states (DOS) of \aFeGe ($0.45 \leq x \leq 0.61$)and a-Fe$_{0.65}$Si$_{0.35}$~\cite{Karel2014a} shown in Fig.~\ref{fDoS} for a more complete explanation of the role of Ge and Si atomic size. 
Intuitively, substituting the larger, isoelectronic Ge atom for Si in a \emph{crystalline} system will increase the lattice constant, shrink the Brillouin zone, and compress the energy bands in $\vec{k}$-space, thereby narrowing and sharpening the features in the density of states . 
For our \emph{amorphous} systems, $\vec{k}$ is no longer a good quantum number so Brillouin zones and energy bands are not meaningful; however, the average interatomic spacing and the density of states make no reference to $\vec{k}$ and so provide a meaningful extension of our crystalline intuition into amorphous systems.
The reduced atomic number density corresponds to an increased average interatomic spacing in \aFeGe compared to \aFeSi. 
This leads to a sharper DOS peak and a stronger spin split near the Fermi level, as shown in Fig.~\ref{fDoS}, which explains the enhanced magnetization, as well as its pronounced $x$-dependence, in \aFeGe when considered in the framework of a simple Stoner band model of itinerant ferromagnetism.~\cite{Stoner1938} We therefore conclude that the scattering from the disorder associated with the amorphous structure is sufficiently weak that it does not substantially redistribute the majority and minority DOS of the Fe-$3d$ electrons,  so the exchange interaction remains strong enough to spin-split the density of states and yield a net magnetization.

Furthermore, the density of states in Fig.~\ref{fDoS} allows us to approximate the spin polarization in \aFeGe; for example, we count states at the Fermi energy $\varepsilon_F$ for \textit{a}-Fe$_{0.61}$Ge$_{0.39}$ and estimate a 37\% spin polarization, nearly identical to the calculated spin polarization for \textit{a}-Fe$_{0.65}$Si$_{0.35}$.~\cite{Karel2014a} 
The low-temperature spin polarization of \aFeSi with $0.58 \leq x \leq 0.68$ has been measured using point-contact Andreev reflection spectroscopy and was found to peak at almost 70\% for $x = 0.65$,~\cite{Karel2018} roughly double the prediction from the calculated DOS; this discrepancy was there attributed to the small size of the supercell used in the calculations. The result in \aFeSi  leads us to suspect a similarly large spin polarization exists in \aFeGe.

\subsection{Anomalous Hall Effect}
These films fall at the high conductivity edge of the low conductivity regime, with $10^3 < \sigma_{xx} < 10^4 \,(\mathrm{\Omega\; cm})^{-1}$ and finite $\rho_{xx}$ as $T \to 0$, so hopping conduction does not apply. To apply a standard empirical scaling argument to the AHC in \aFeGe, an appropriate normalization is essential; we consider two factors to enable comparison of our results to the scaling of the AHC in \aFeSi.~\cite{Karel2016}
First, in changing the Fe concentration $x$, $M_z$ is also modified. 
Since $\sigma_{xy}^{\mathrm{AH}} \propto \rho_{xy}^{\mathrm{AH}} \propto M_z$, changing $x$ alters $\sigma_{xy}^{\mathrm{AH}}$ directly due to the dependence on $M_z$; we remove this dependence by dividing $\sigma_{xy}^{\mathrm{AH}}$ by $M_z$ and obtain the data shown in Fig.~\ref{f4}(a).
These data can be fit by $\sigma_{xy}^{\mathrm{AH}} \propto \sigma_{xx}^{1.1}$, similar to what was obtained for \aFeSi but inconsistent with the theoretical predictions of Onoda~\cite{Onoda2006} and Liu.~\cite{Liu2011} 
Additionally, since the unified theory is expressed in terms of the carrier lifetime $\tau$ and not the conductivity $\sigma_{xx}$, we must account for the effect of changing $x$ on $\tau$. Hence, we consider a simple free-electron-type model which gives $n_h^{2/3}$ as our second normalization factor, using the values of $n_h$ shown in Fig.~\ref{fRho}(b). This normalization yields the scaling plot in Fig.~\ref{f4}(b), in which $\sigma_{xy}^{\mathrm{AH}}/M_z n_h^{2/3}$ is essentially independent of $\sigma_{xx}$. Both scaling plots in Fig.~\ref{f4} also show the appropriately normalized anomalous Hall conductivity in \aFeSi from Ref.~\citenum{Karel2016} as a point of comparison; we attribute the quantitative similarity in the normalized AHC of \aFeGe and \aFeSi to the spin-orbit coupling introduced near $\varepsilon_F$ by the Fe $d$-states in both systems, since the spin-orbit coupling due to the Ge (or Si) $p$-states is diluted over a wider energy range.

\begin{figure}
    \includegraphics{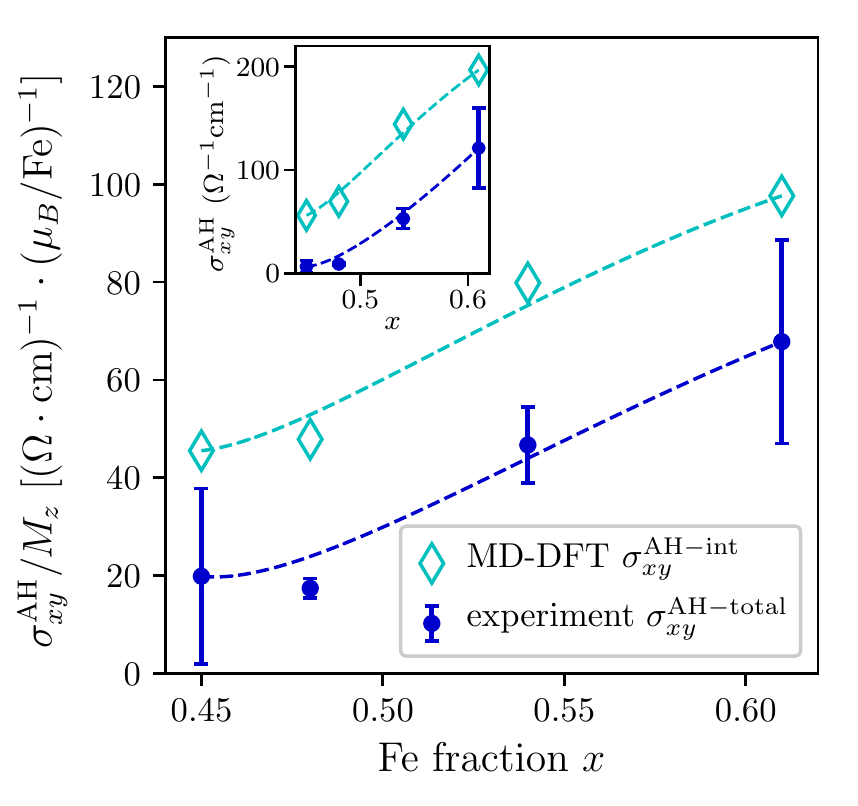}
    \caption{Comparison of calculated \emph{intrinsic} anomalous Hall conductivity from 128-atom MD-DFT (turquoise diamonds) to experimentally measured \emph{total} anomalous Hall conductivity (blue dots). The main axis shows the AHC normalized by the saturation magnetization $M_z$, where experimental values have been normalized by the experimentally measured $M_z$ and theoretical values have been normalized by the theoretically calculated $M_z$ for consistency. Unnormalized values of the measured total and calculated intrinsic AHC are shown in the inset. Spline fits to data (dashed lines) are shown as a guide to the eye.}
    \label{f:AHCtheoryexpt}
\end{figure}

The fact that $\sigma_{xy}^{\mathrm{AH}}/M_z n_h^{2/3}$ is independent of $\sigma_{xx}$ implies that either the side-jump or the intrinsic mechanism, or some combination of the two, is responsible for the anomalous Hall conductivity. 
In the case of \aFeSi, these two contributions were qualitatively deconvolved by comparison to measurements on crystalline Fe$_x$Si$_{1-x}$, which were compared to calculations of the intrinsic and side-jump mechanisms in the AHC of bcc Fe.~\cite{Karel2016} 
To more rigorously separate the contributions of these two mechanisms in \aFeGe, we compare \textit{ab initio} values of the \textit{intrinsic} anomalous Hall conductivity calculated from our spin-orbit-coupled 128-atom MD-DFT, which are normalized by the calculated magnetization and shown in Fig.~\ref{f:AHCtheoryexpt}, to experimental values of the \textit{total} anomalous Hall conductivity, normalized by the measured magnetization; this comparison shows excellent agreement in the trend of the calculated and experimental values, suggesting that the measured AHC consists of a large intrinsic component, which varies with composition and is partially offset by a composition-independent side-jump component. 
While the intrinsic AHC is generally expressed as the integral of the Berry curvature taken over all occupied bands, this result independently verifies the outcome of our scaling argument and unambiguously shows that the intrinsic mechanism persists in systems whose band structure is ill-defined, further validating the real-space formulation of the AHC in Ref.~\citenum{Marrazzo2017a}. 


\begin{figure}
    \centering
    \includegraphics{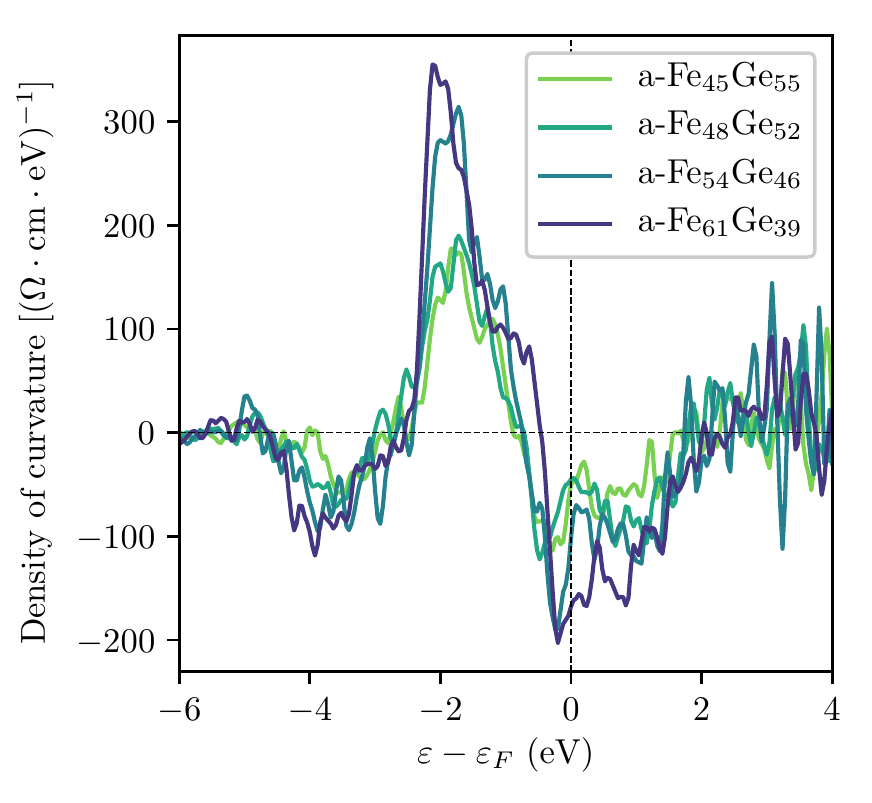}
    \caption{Density of Berry curvature as a function of chemical potential calculated from 128-atom MD-DFT for \aFeGe with $x = 0.45$, 0.48, 0.54 and 0.61. }
    \label{f:DOC}
\end{figure}

Our spin-orbit coupled DFT calculations provide computational foundations and a theoretical paradigm shift for understanding the \emph{intrinsic} anomalous Hall conductivity in an amorphous material. The intrinsic AHC is the integral over occupied states of the Berry curvature; however, as $\vec{k}$ is not a good quantum number in an amorphous system, expressing the Berry curvature as $\Omega(\vec{k})$ is meaningless here.
Instead, we construct a framework appropriate to our system based on the \textit{density of Berry curvature}:~\cite{Guo2008, Sahin2015}
\begin{equation}
    \rho_{\mathrm{DOC}}(\varepsilon) = \sum_{\vec{k}}^{} \Omega(\vec{k}) \delta(\varepsilon_{\vec{k}} - \varepsilon)
\end{equation}
which corresponds to the Berry curvature within an energy range bounded by $\varepsilon$ and $\varepsilon + d\varepsilon$ integrated throughout the Brillouin zone. 
This can, in principle, be computed in an amorphous material without reference to $\vec{k}$-space by adding together the partial densities of states for local orbital states with spin-orbit correlation parallel and antiparallel, which are energetically offset by the self-consistently determined exchange energy.
If the energy scales of the disorder potentials that scatter states from momentum $\vec{k}$ to $\vec{k}'$ are relatively weak compared to the energy scale spanned by the features in $\rho_{\mathrm{DOC}}(\varepsilon)$, then the density of Berry curvature would be approximately preserved even as momentum fails to remain a good quantum number in an amorphous system.

Fig.~\ref{f:DOC} shows the density of Berry curvature as a function of chemical potential for the compositions $x$ of \aFeGe studied in this work; integrating the density of curvature over occupied energy states yields the intrinsic AHC shown in Fig.~\ref{f:AHCtheoryexpt}, in the same way that integrating the spin-resolved density of states over occupied energy states yields the magnetization.
Fig.~\ref{f:DOC} further indicates that the energy scale of the density of Berry curvature is on the order of 5 eV, an energy scale set in part by the spin-orbit interaction-induced splitting of parallel and antiparallel spin-orbit correlated states and in part by the exchange energy in our system. 
As long as the disorder potentials are typically smaller than 5 eV our conclusions about the AHC would remain similar in both the clean and disordered systems; otherwise, the spin-orbit correlations that give rise to the AHC would be lost.
This energy scale lends confidence that the calculated density of curvature, which imposes an artificial periodicity on the simulated 128-atom amorphous supercell, is a suitable approximation for that of the fully amorphous material: the similarity between the simulated and measured atomic number densities [Fig.~\ref{f2}(b)] indicates that the supercell accurately reproduces the structural disorder, and hence the disorder potentials, of the actual amorphous structure.

\section{Conclusion}
In summary, we presented experimental and computational studies of the magnetic and transport properties of amorphous Fe$_x$Ge$_{1-x}$ ($0.45 \leq x \leq 0.61$) thin films, including the first measurement of the anomalous Hall effect in \aFeGe. 
Its magnetization is well explained by a Stoner band model above the onset of magnetic order around $x=0.4$, with the larger Ge atom distorting the density of states and causing enhanced magnetization at all $x$ in \aFeGe compared to \aFeSi.
The resistivity of \aFeGe agrees with established work on amorphous metals, and indicates that a metal-insulator transition occurs at lower $x$ than the onset of ferromagnetism. 
The anomalous Hall conductivity of \aFeGe is independent of the longitudinal conductivity when appropriately normalized by $M_z n_h^{2/3}$ to account for the role of changing the Fe concentration on the magnetization and carrier lifetime; our DFT calculations refine this scaling argument and indicate the AHC is comprised of a dominant intrinsic component whose magnitude is influenced by the Fe concentration, and of an opposing composition-independent side-jump component.
The calculated density of Berry curvature shows that the AHC in this amorphous material is indeed intrinsic; moreover, because the density of curvature can be calculated from either the spin-orbit correlations of local orbital states in a disordered material or the band structure in a crystalline one, it provides a novel and versatile Stoner-esque model for understanding the electronic origins of the intrinsic AHC in systems possessing and lacking long-range order.
Since the spin and orbital Hall conductivities arise from mathematically analogous Berry curvature-like quantities, we anticipate that the straightforward extension of this model to those phenomena will spur experimental investigation of the spin and orbital Hall effects in amorphous materials, portending future application of such materials as spin and orbital torque generators in next-generation devices.

\begin{acknowledgements}
The authors thank C. \c{S}ahin, J. Karel, S. Mack, and N. D. Reynolds for illuminating discussions.
This work was primarily funded by the U.S. Department of Energy, Office of Science, Office of Basic Energy Sciences, Materials Sciences and Engineering Division under Contract No. DE-AC02-05-CH11231 (NEMM program MSMAG). 
Work for 128-atom DFT calculations was supported by DOE-BES (Grant No. DE-FG02-05-ER46237). Additionally, BHZ acknowledges support from the Basic Research Program of China (2015CB921400).
Support for the density of Berry curvature framework for the AHC was provided by the Center for Emergent Materials, an NSF MRSEC under Award No. DMR-1420451.
\end{acknowledgements}


\begin{thebibliography}{36}%
\makeatletter
\providecommand \@ifxundefined [1]{%
 \@ifx{#1\undefined}
}%
\providecommand \@ifnum [1]{%
 \ifnum #1\expandafter \@firstoftwo
 \else \expandafter \@secondoftwo
 \fi
}%
\providecommand \@ifx [1]{%
 \ifx #1\expandafter \@firstoftwo
 \else \expandafter \@secondoftwo
 \fi
}%
\providecommand \natexlab [1]{#1}%
\providecommand \enquote  [1]{``#1''}%
\providecommand \bibnamefont  [1]{#1}%
\providecommand \bibfnamefont [1]{#1}%
\providecommand \citenamefont [1]{#1}%
\providecommand \href@noop [0]{\@secondoftwo}%
\providecommand \href [0]{\begingroup \@sanitize@url \@href}%
\providecommand \@href[1]{\@@startlink{#1}\@@href}%
\providecommand \@@href[1]{\endgroup#1\@@endlink}%
\providecommand \@sanitize@url [0]{\catcode `\\12\catcode `\$12\catcode
  `\&12\catcode `\#12\catcode `\^12\catcode `\_12\catcode `\%12\relax}%
\providecommand \@@startlink[1]{}%
\providecommand \@@endlink[0]{}%
\providecommand \url  [0]{\begingroup\@sanitize@url \@url }%
\providecommand \@url [1]{\endgroup\@href {#1}{\urlprefix }}%
\providecommand \urlprefix  [0]{URL }%
\providecommand \Eprint [0]{\href }%
\providecommand \doibase [0]{http://dx.doi.org/}%
\providecommand \selectlanguage [0]{\@gobble}%
\providecommand \bibinfo  [0]{\@secondoftwo}%
\providecommand \bibfield  [0]{\@secondoftwo}%
\providecommand \translation [1]{[#1]}%
\providecommand \BibitemOpen [0]{}%
\providecommand \bibitemStop [0]{}%
\providecommand \bibitemNoStop [0]{.\EOS\space}%
\providecommand \EOS [0]{\spacefactor3000\relax}%
\providecommand \BibitemShut  [1]{\csname bibitem#1\endcsname}%
\let\auto@bib@innerbib\@empty
\bibitem [{\citenamefont {Nagaosa}\ \emph {et~al.}(2010)\citenamefont
  {Nagaosa}, \citenamefont {Sinova}, \citenamefont {Onoda}, \citenamefont
  {MacDonald},\ and\ \citenamefont {Ong}}]{Nagaosa2010}%
  \BibitemOpen
  \bibfield  {author} {\bibinfo {author} {\bibfnamefont {N.}~\bibnamefont
  {Nagaosa}}, \bibinfo {author} {\bibfnamefont {J.}~\bibnamefont {Sinova}},
  \bibinfo {author} {\bibfnamefont {S.}~\bibnamefont {Onoda}}, \bibinfo
  {author} {\bibfnamefont {A.~H.}\ \bibnamefont {MacDonald}}, \ and\ \bibinfo
  {author} {\bibfnamefont {N.~P.}\ \bibnamefont {Ong}},\ }\href {\doibase
  10.1103/RevModPhys.82.1539} {\bibfield  {journal} {\bibinfo  {journal} {Rev.
  Mod. Phys.}\ }\textbf {\bibinfo {volume} {82}},\ \bibinfo {pages} {1539}
  (\bibinfo {year} {2010})}\BibitemShut {NoStop}%
\bibitem [{\citenamefont {Sinova}\ \emph {et~al.}(2015)\citenamefont {Sinova},
  \citenamefont {Valenzuela}, \citenamefont {Wunderlich}, \citenamefont
  {Back},\ and\ \citenamefont {Jungwirth}}]{Sinova2015}%
  \BibitemOpen
  \bibfield  {author} {\bibinfo {author} {\bibfnamefont {J.}~\bibnamefont
  {Sinova}}, \bibinfo {author} {\bibfnamefont {S.~O.}\ \bibnamefont
  {Valenzuela}}, \bibinfo {author} {\bibfnamefont {J.}~\bibnamefont
  {Wunderlich}}, \bibinfo {author} {\bibfnamefont {C.~H.}\ \bibnamefont
  {Back}}, \ and\ \bibinfo {author} {\bibfnamefont {T.}~\bibnamefont
  {Jungwirth}},\ }\href {\doibase 10.1103/RevModPhys.87.1213} {\bibfield
  {journal} {\bibinfo  {journal} {Rev. Mod. Phys.}\ }\textbf {\bibinfo {volume}
  {87}},\ \bibinfo {pages} {1213} (\bibinfo {year} {2015})}\BibitemShut
  {NoStop}%
\bibitem [{Note1()}]{Note1}%
  \BibitemOpen
  \bibinfo {note} {While the AHE was long understood to be present exclusively
  in ferromagnets, recent results indicate that certain noncollinear
  antiferromagnets also host a substantial AHE~[A. K. Nayak \protect \emph {et
  al.}, Sci. Adv. \protect \textbf {2}, e1501870 (2016)].}\BibitemShut {Stop}%
\bibitem [{\citenamefont {Pugh}(1930)}]{Pugh1930}%
  \BibitemOpen
  \bibfield  {author} {\bibinfo {author} {\bibfnamefont {E.~M.}\ \bibnamefont
  {Pugh}},\ }\href {\doibase 10.1103/PhysRev.36.1503} {\bibfield  {journal}
  {\bibinfo  {journal} {Phys. Rev.}\ }\textbf {\bibinfo {volume} {36}},\
  \bibinfo {pages} {1503} (\bibinfo {year} {1930})}\BibitemShut {NoStop}%
\bibitem [{\citenamefont {Smit}(1955)}]{Smit1955}%
  \BibitemOpen
  \bibfield  {author} {\bibinfo {author} {\bibfnamefont {J.}~\bibnamefont
  {Smit}},\ }\href {\doibase 10.1016/S0031-8914(55)92596-9} {\bibfield
  {journal} {\bibinfo  {journal} {Physica}\ }\textbf {\bibinfo {volume} {21}},\
  \bibinfo {pages} {877} (\bibinfo {year} {1955})}\BibitemShut {NoStop}%
\bibitem [{\citenamefont {Berger}(1970)}]{Berger1970}%
  \BibitemOpen
  \bibfield  {author} {\bibinfo {author} {\bibfnamefont {L.}~\bibnamefont
  {Berger}},\ }\href {\doibase 10.1103/PhysRevB.2.4559} {\bibfield  {journal}
  {\bibinfo  {journal} {Phys. Rev. B}\ }\textbf {\bibinfo {volume} {2}},\
  \bibinfo {pages} {4559} (\bibinfo {year} {1970})}\BibitemShut {NoStop}%
\bibitem [{\citenamefont {Xiao}\ \emph {et~al.}(2010)\citenamefont {Xiao},
  \citenamefont {Chang},\ and\ \citenamefont {Niu}}]{Xiao2010}%
  \BibitemOpen
  \bibfield  {author} {\bibinfo {author} {\bibfnamefont {D.}~\bibnamefont
  {Xiao}}, \bibinfo {author} {\bibfnamefont {M.-C.~C.}\ \bibnamefont {Chang}},
  \ and\ \bibinfo {author} {\bibfnamefont {Q.}~\bibnamefont {Niu}},\ }\href
  {\doibase 10.1103/RevModPhys.82.1959} {\bibfield  {journal} {\bibinfo
  {journal} {Rev. Mod. Phys.}\ }\textbf {\bibinfo {volume} {82}},\ \bibinfo
  {pages} {1959} (\bibinfo {year} {2010})}\BibitemShut {NoStop}%
\bibitem [{\citenamefont {Jungwirth}\ \emph {et~al.}(2002)\citenamefont
  {Jungwirth}, \citenamefont {Niu},\ and\ \citenamefont
  {Macdonald}}]{Jungwirth2002}%
  \BibitemOpen
  \bibfield  {author} {\bibinfo {author} {\bibfnamefont {T.}~\bibnamefont
  {Jungwirth}}, \bibinfo {author} {\bibfnamefont {Q.}~\bibnamefont {Niu}}, \
  and\ \bibinfo {author} {\bibfnamefont {A.~H.}\ \bibnamefont {Macdonald}},\
  }\href {\doibase 10.1103/PhysRevLett.88.207208} {\bibfield  {journal}
  {\bibinfo  {journal} {Phys. Rev. Lett.}\ }\textbf {\bibinfo {volume} {88}},\
  \bibinfo {pages} {207208} (\bibinfo {year} {2002})}\BibitemShut {NoStop}%
\bibitem [{\citenamefont {Onoda}\ and\ \citenamefont
  {Nagaosa}(2002)}]{Onoda2002}%
  \BibitemOpen
  \bibfield  {author} {\bibinfo {author} {\bibfnamefont {M.}~\bibnamefont
  {Onoda}}\ and\ \bibinfo {author} {\bibfnamefont {N.}~\bibnamefont
  {Nagaosa}},\ }\href {\doibase 10.1143/JPSJ.71.19} {\bibfield  {journal}
  {\bibinfo  {journal} {J. Phys. Soc. Japan}\ }\textbf {\bibinfo {volume}
  {71}},\ \bibinfo {pages} {19} (\bibinfo {year} {2002})}\BibitemShut {NoStop}%
\bibitem [{\citenamefont {Onoda}\ \emph {et~al.}(2006)\citenamefont {Onoda},
  \citenamefont {Sugimoto},\ and\ \citenamefont {Nagaosa}}]{Onoda2006}%
  \BibitemOpen
  \bibfield  {author} {\bibinfo {author} {\bibfnamefont {S.}~\bibnamefont
  {Onoda}}, \bibinfo {author} {\bibfnamefont {N.}~\bibnamefont {Sugimoto}}, \
  and\ \bibinfo {author} {\bibfnamefont {N.}~\bibnamefont {Nagaosa}},\ }\href
  {\doibase 10.1103/PhysRevLett.97.126602} {\bibfield  {journal} {\bibinfo
  {journal} {Phys. Rev. Lett.}\ }\textbf {\bibinfo {volume} {97}},\ \bibinfo
  {pages} {126602} (\bibinfo {year} {2006})}\BibitemShut {NoStop}%
\bibitem [{\citenamefont {Liu}\ \emph {et~al.}(2011)\citenamefont {Liu},
  \citenamefont {Liu},\ and\ \citenamefont {Sinova}}]{Liu2011}%
  \BibitemOpen
  \bibfield  {author} {\bibinfo {author} {\bibfnamefont {X.-J.}\ \bibnamefont
  {Liu}}, \bibinfo {author} {\bibfnamefont {X.}~\bibnamefont {Liu}}, \ and\
  \bibinfo {author} {\bibfnamefont {J.}~\bibnamefont {Sinova}},\ }\href
  {\doibase 10.1103/PhysRevB.84.165304} {\bibfield  {journal} {\bibinfo
  {journal} {Phys. Rev. B}\ }\textbf {\bibinfo {volume} {84}},\ \bibinfo
  {pages} {165304} (\bibinfo {year} {2011})}\BibitemShut {NoStop}%
\bibitem [{\citenamefont {Karel}\ \emph {et~al.}(2016)\citenamefont {Karel},
  \citenamefont {Bordel}, \citenamefont {Bouma}, \citenamefont
  {de~Lorimier-Farmer}, \citenamefont {Lee},\ and\ \citenamefont
  {Hellman}}]{Karel2016}%
  \BibitemOpen
  \bibfield  {author} {\bibinfo {author} {\bibfnamefont {J.}~\bibnamefont
  {Karel}}, \bibinfo {author} {\bibfnamefont {C.}~\bibnamefont {Bordel}},
  \bibinfo {author} {\bibfnamefont {D.~S.}\ \bibnamefont {Bouma}}, \bibinfo
  {author} {\bibfnamefont {A.}~\bibnamefont {de~Lorimier-Farmer}}, \bibinfo
  {author} {\bibfnamefont {H.~J.}\ \bibnamefont {Lee}}, \ and\ \bibinfo
  {author} {\bibfnamefont {F.}~\bibnamefont {Hellman}},\ }\href {\doibase
  10.1209/0295-5075/114/57004} {\bibfield  {journal} {\bibinfo  {journal} {EPL
  (Europhysics Lett.)}\ }\textbf {\bibinfo {volume} {114}},\ \bibinfo {pages}
  {57004} (\bibinfo {year} {2016})}\BibitemShut {NoStop}%
\bibitem [{\citenamefont {{\c{S}}ahin}\ and\ \citenamefont
  {Flatt{\'{e}}}(2015)}]{Sahin2015}%
  \BibitemOpen
  \bibfield  {author} {\bibinfo {author} {\bibfnamefont {C.}~\bibnamefont
  {{\c{S}}ahin}}\ and\ \bibinfo {author} {\bibfnamefont {M.~E.}\ \bibnamefont
  {Flatt{\'{e}}}},\ }\href {\doibase 10.1103/PhysRevLett.114.107201} {\bibfield
   {journal} {\bibinfo  {journal} {Phys. Rev. Lett.}\ }\textbf {\bibinfo
  {volume} {114}},\ \bibinfo {pages} {107201} (\bibinfo {year}
  {2015})}\BibitemShut {NoStop}%
\bibitem [{\citenamefont {Guo}\ \emph {et~al.}(2008)\citenamefont {Guo},
  \citenamefont {Murakami}, \citenamefont {Chen},\ and\ \citenamefont
  {Nagaosa}}]{Guo2008}%
  \BibitemOpen
  \bibfield  {author} {\bibinfo {author} {\bibfnamefont {G.~Y.}\ \bibnamefont
  {Guo}}, \bibinfo {author} {\bibfnamefont {S.}~\bibnamefont {Murakami}},
  \bibinfo {author} {\bibfnamefont {T.-W.}\ \bibnamefont {Chen}}, \ and\
  \bibinfo {author} {\bibfnamefont {N.}~\bibnamefont {Nagaosa}},\ }\href
  {\doibase 10.1103/PhysRevLett.100.096401} {\bibfield  {journal} {\bibinfo
  {journal} {Phys. Rev. Lett.}\ }\textbf {\bibinfo {volume} {100}},\ \bibinfo
  {pages} {096401} (\bibinfo {year} {2008})}\BibitemShut {NoStop}%
\bibitem [{\citenamefont {Marrazzo}\ and\ \citenamefont
  {Resta}(2017)}]{Marrazzo2017a}%
  \BibitemOpen
  \bibfield  {author} {\bibinfo {author} {\bibfnamefont {A.}~\bibnamefont
  {Marrazzo}}\ and\ \bibinfo {author} {\bibfnamefont {R.}~\bibnamefont
  {Resta}},\ }\href {\doibase 10.1103/PhysRevB.95.121114} {\bibfield  {journal}
  {\bibinfo  {journal} {Phys. Rev. B}\ }\textbf {\bibinfo {volume} {95}},\
  \bibinfo {pages} {121114(R)} (\bibinfo {year} {2017})}\BibitemShut {NoStop}%
\bibitem [{\citenamefont {Huang}\ and\ \citenamefont
  {Chien}(2012)}]{Huang2012}%
  \BibitemOpen
  \bibfield  {author} {\bibinfo {author} {\bibfnamefont {S.~X.}\ \bibnamefont
  {Huang}}\ and\ \bibinfo {author} {\bibfnamefont {C.~L.}\ \bibnamefont
  {Chien}},\ }\href {\doibase 10.1103/PhysRevLett.108.267201} {\bibfield
  {journal} {\bibinfo  {journal} {Phys. Rev. Lett.}\ }\textbf {\bibinfo
  {volume} {108}},\ \bibinfo {pages} {267201} (\bibinfo {year}
  {2012})}\BibitemShut {NoStop}%
\bibitem [{\citenamefont {Porter}\ \emph {et~al.}(2014)\citenamefont {Porter},
  \citenamefont {Gartside},\ and\ \citenamefont {Marrows}}]{Porter2014}%
  \BibitemOpen
  \bibfield  {author} {\bibinfo {author} {\bibfnamefont {N.~A.}\ \bibnamefont
  {Porter}}, \bibinfo {author} {\bibfnamefont {J.~C.}\ \bibnamefont
  {Gartside}}, \ and\ \bibinfo {author} {\bibfnamefont {C.~H.}\ \bibnamefont
  {Marrows}},\ }\href {\doibase 10.1103/PhysRevB.90.024403} {\bibfield
  {journal} {\bibinfo  {journal} {Phys. Rev. B}\ }\textbf {\bibinfo {volume}
  {90}},\ \bibinfo {pages} {024403} (\bibinfo {year} {2014})}\BibitemShut
  {NoStop}%
\bibitem [{\citenamefont {Gallagher}\ \emph {et~al.}(2017)\citenamefont
  {Gallagher}, \citenamefont {Meng}, \citenamefont {Brangham}, \citenamefont
  {Wang}, \citenamefont {Esser}, \citenamefont {McComb},\ and\ \citenamefont
  {Yang}}]{Gallagher2016a}%
  \BibitemOpen
  \bibfield  {author} {\bibinfo {author} {\bibfnamefont {J.~C.}\ \bibnamefont
  {Gallagher}}, \bibinfo {author} {\bibfnamefont {K.~Y.}\ \bibnamefont {Meng}},
  \bibinfo {author} {\bibfnamefont {J.~T.}\ \bibnamefont {Brangham}}, \bibinfo
  {author} {\bibfnamefont {H.~L.}\ \bibnamefont {Wang}}, \bibinfo {author}
  {\bibfnamefont {B.~D.}\ \bibnamefont {Esser}}, \bibinfo {author}
  {\bibfnamefont {D.~W.}\ \bibnamefont {McComb}}, \ and\ \bibinfo {author}
  {\bibfnamefont {F.~Y.}\ \bibnamefont {Yang}},\ }\href {\doibase
  10.1103/PhysRevLett.118.027201} {\bibfield  {journal} {\bibinfo  {journal}
  {Phys. Rev. Lett.}\ }\textbf {\bibinfo {volume} {118}},\ \bibinfo {pages}
  {027201} (\bibinfo {year} {2017})}\BibitemShut {NoStop}%
\bibitem [{\citenamefont {Suran}\ \emph {et~al.}(1976)\citenamefont {Suran},
  \citenamefont {Daver},\ and\ \citenamefont {Bruyere}}]{Suran1976}%
  \BibitemOpen
  \bibfield  {author} {\bibinfo {author} {\bibfnamefont {G.}~\bibnamefont
  {Suran}}, \bibinfo {author} {\bibfnamefont {H.}~\bibnamefont {Daver}}, \ and\
  \bibinfo {author} {\bibfnamefont {J.~C.}\ \bibnamefont {Bruyere}},\ }in\
  \href {\doibase 10.1063/1.30562} {\emph {\bibinfo {booktitle} {AIP Conf.
  Proc.}}},\ Vol.~\bibinfo {volume} {29}\ (\bibinfo  {publisher} {AIP},\
  \bibinfo {year} {1976})\ pp.\ \bibinfo {pages} {162--164}\BibitemShut
  {NoStop}%
\bibitem [{\citenamefont {Daver}\ and\ \citenamefont
  {Massenet}(1977)}]{Daver1977}%
  \BibitemOpen
  \bibfield  {author} {\bibinfo {author} {\bibfnamefont {H.}~\bibnamefont
  {Daver}}\ and\ \bibinfo {author} {\bibfnamefont {O.}~\bibnamefont
  {Massenet}},\ }\href {\doibase 10.1016/0038-1098(77)90240-X} {\bibfield
  {journal} {\bibinfo  {journal} {Solid State Commun.}\ }\textbf {\bibinfo
  {volume} {23}},\ \bibinfo {pages} {393} (\bibinfo {year} {1977})}\BibitemShut
  {NoStop}%
\bibitem [{\citenamefont {Terzieff}\ \emph {et~al.}(1979)\citenamefont
  {Terzieff}, \citenamefont {Lee},\ and\ \citenamefont
  {Heiman}}]{Terzieff1979}%
  \BibitemOpen
  \bibfield  {author} {\bibinfo {author} {\bibfnamefont {P.}~\bibnamefont
  {Terzieff}}, \bibinfo {author} {\bibfnamefont {K.}~\bibnamefont {Lee}}, \
  and\ \bibinfo {author} {\bibfnamefont {N.}~\bibnamefont {Heiman}},\ }\href
  {\doibase 10.1063/1.326101} {\bibfield  {journal} {\bibinfo  {journal} {J.
  Appl. Phys.}\ }\textbf {\bibinfo {volume} {50}},\ \bibinfo {pages} {1031}
  (\bibinfo {year} {1979})}\BibitemShut {NoStop}%
\bibitem [{\citenamefont {Randhawa}\ \emph {et~al.}(1981)\citenamefont
  {Randhawa}, \citenamefont {Malhotra},\ and\ \citenamefont
  {Chopra}}]{Randhawa1981}%
  \BibitemOpen
  \bibfield  {author} {\bibinfo {author} {\bibfnamefont {H.}~\bibnamefont
  {Randhawa}}, \bibinfo {author} {\bibfnamefont {L.~K.}\ \bibnamefont
  {Malhotra}}, \ and\ \bibinfo {author} {\bibfnamefont {K.~L.}\ \bibnamefont
  {Chopra}},\ }\href {\doibase 10.1063/1.329649} {\bibfield  {journal}
  {\bibinfo  {journal} {J. Appl. Phys.}\ }\textbf {\bibinfo {volume} {52}},\
  \bibinfo {pages} {1600} (\bibinfo {year} {1981})}\BibitemShut {NoStop}%
\bibitem [{\citenamefont {Lorentz}\ \emph {et~al.}(1984)\citenamefont
  {Lorentz}, \citenamefont {Laderman},\ and\ \citenamefont
  {Bienenstock}}]{Lorentz1984}%
  \BibitemOpen
  \bibfield  {author} {\bibinfo {author} {\bibfnamefont {R.~D.}\ \bibnamefont
  {Lorentz}}, \bibinfo {author} {\bibfnamefont {S.~S.}\ \bibnamefont
  {Laderman}}, \ and\ \bibinfo {author} {\bibfnamefont {A.~I.}\ \bibnamefont
  {Bienenstock}},\ }in\ \href {\doibase 10.1007/978-3-642-46522-2_70} {\emph
  {\bibinfo {booktitle} {EXAFS Near Edge Struct. III}}}\ (\bibinfo {year}
  {1984})\ pp.\ \bibinfo {pages} {280--283}\BibitemShut {NoStop}%
\bibitem [{\citenamefont {Gray}\ \emph {et~al.}(2011)\citenamefont {Gray},
  \citenamefont {Karel}, \citenamefont {Min{\'{a}}r}, \citenamefont {Bordel},
  \citenamefont {Ebert}, \citenamefont {Braun}, \citenamefont {Ueda},
  \citenamefont {Yamashita}, \citenamefont {Ouyang}, \citenamefont {Smith},
  \citenamefont {Kobayashi}, \citenamefont {Hellman},\ and\ \citenamefont
  {Fadley}}]{Gray2011}%
  \BibitemOpen
  \bibfield  {author} {\bibinfo {author} {\bibfnamefont {A.~X.}\ \bibnamefont
  {Gray}}, \bibinfo {author} {\bibfnamefont {J.}~\bibnamefont {Karel}},
  \bibinfo {author} {\bibfnamefont {J.}~\bibnamefont {Min{\'{a}}r}}, \bibinfo
  {author} {\bibfnamefont {C.}~\bibnamefont {Bordel}}, \bibinfo {author}
  {\bibfnamefont {H.}~\bibnamefont {Ebert}}, \bibinfo {author} {\bibfnamefont
  {J.}~\bibnamefont {Braun}}, \bibinfo {author} {\bibfnamefont
  {S.}~\bibnamefont {Ueda}}, \bibinfo {author} {\bibfnamefont {Y.}~\bibnamefont
  {Yamashita}}, \bibinfo {author} {\bibfnamefont {L.}~\bibnamefont {Ouyang}},
  \bibinfo {author} {\bibfnamefont {D.~J.}\ \bibnamefont {Smith}}, \bibinfo
  {author} {\bibfnamefont {K.}~\bibnamefont {Kobayashi}}, \bibinfo {author}
  {\bibfnamefont {F.}~\bibnamefont {Hellman}}, \ and\ \bibinfo {author}
  {\bibfnamefont {C.~S.}\ \bibnamefont {Fadley}},\ }\href {\doibase
  10.1103/PhysRevB.83.195112} {\bibfield  {journal} {\bibinfo  {journal} {Phys.
  Rev. B}\ }\textbf {\bibinfo {volume} {83}},\ \bibinfo {pages} {195112}
  (\bibinfo {year} {2011})}\BibitemShut {NoStop}%
\bibitem [{\citenamefont {van~der Pauw}(1958)}]{VanderPauw1958}%
  \BibitemOpen
  \bibfield  {author} {\bibinfo {author} {\bibfnamefont {L.~J.}\ \bibnamefont
  {van~der Pauw}},\ }\href@noop {} {\bibfield  {journal} {\bibinfo  {journal}
  {Philips Res. Reports}\ }\textbf {\bibinfo {volume} {13}},\ \bibinfo {pages}
  {1} (\bibinfo {year} {1958})}\BibitemShut {NoStop}%
\bibitem [{\citenamefont {Bl{\"{o}}chl}(1994)}]{Blochl1994}%
  \BibitemOpen
  \bibfield  {author} {\bibinfo {author} {\bibfnamefont {P.~E.}\ \bibnamefont
  {Bl{\"{o}}chl}},\ }\href {\doibase 10.1103/PhysRevB.50.17953} {\bibfield
  {journal} {\bibinfo  {journal} {Phys. Rev. B}\ }\textbf {\bibinfo {volume}
  {50}},\ \bibinfo {pages} {17953} (\bibinfo {year} {1994})}\BibitemShut
  {NoStop}%
\bibitem [{\citenamefont {Kresse}\ and\ \citenamefont
  {Joubert}(1999)}]{Kresse1999}%
  \BibitemOpen
  \bibfield  {author} {\bibinfo {author} {\bibfnamefont {G.}~\bibnamefont
  {Kresse}}\ and\ \bibinfo {author} {\bibfnamefont {D.}~\bibnamefont
  {Joubert}},\ }\href {\doibase 10.1103/PhysRevB.59.1758} {\bibfield  {journal}
  {\bibinfo  {journal} {Phys. Rev. B}\ }\textbf {\bibinfo {volume} {59}},\
  \bibinfo {pages} {1758} (\bibinfo {year} {1999})}\BibitemShut {NoStop}%
\bibitem [{\citenamefont {Kresse}\ and\ \citenamefont
  {Hafner}(1993)}]{Kresse1993}%
  \BibitemOpen
  \bibfield  {author} {\bibinfo {author} {\bibfnamefont {G.}~\bibnamefont
  {Kresse}}\ and\ \bibinfo {author} {\bibfnamefont {J.}~\bibnamefont
  {Hafner}},\ }\href {\doibase 10.1103/PhysRevB.47.558} {\bibfield  {journal}
  {\bibinfo  {journal} {Phys. Rev. B}\ }\textbf {\bibinfo {volume} {47}},\
  \bibinfo {pages} {558} (\bibinfo {year} {1993})}\BibitemShut {NoStop}%
\bibitem [{\citenamefont {Kresse}\ and\ \citenamefont
  {Furthm{\"{u}}ller}(1996)}]{Kresse1996}%
  \BibitemOpen
  \bibfield  {author} {\bibinfo {author} {\bibfnamefont {G.}~\bibnamefont
  {Kresse}}\ and\ \bibinfo {author} {\bibfnamefont {J.}~\bibnamefont
  {Furthm{\"{u}}ller}},\ }\href {\doibase 10.1103/PhysRevB.54.11169} {\bibfield
   {journal} {\bibinfo  {journal} {Phys. Rev. B}\ }\textbf {\bibinfo {volume}
  {54}},\ \bibinfo {pages} {11169} (\bibinfo {year} {1996})}\BibitemShut
  {NoStop}%
\bibitem [{\citenamefont {Perdew}\ \emph {et~al.}(1996)\citenamefont {Perdew},
  \citenamefont {Burke},\ and\ \citenamefont {Ernzerhof}}]{Perdew1996}%
  \BibitemOpen
  \bibfield  {author} {\bibinfo {author} {\bibfnamefont {J.~P.}\ \bibnamefont
  {Perdew}}, \bibinfo {author} {\bibfnamefont {K.}~\bibnamefont {Burke}}, \
  and\ \bibinfo {author} {\bibfnamefont {M.}~\bibnamefont {Ernzerhof}},\ }\href
  {\doibase 10.1103/PhysRevLett.77.3865} {\bibfield  {journal} {\bibinfo
  {journal} {Phys. Rev. Lett.}\ }\textbf {\bibinfo {volume} {77}},\ \bibinfo
  {pages} {3865} (\bibinfo {year} {1996})}\BibitemShut {NoStop}%
\bibitem [{\citenamefont {Chen}\ \emph {et~al.}()\citenamefont {Chen},
  \citenamefont {Bouma}, \citenamefont {Kent}, \citenamefont {Streubel},
  \citenamefont {Roy}, \citenamefont {Kevan}, \citenamefont {Fischer},\ and\
  \citenamefont {Hellman}}]{Chen2019}%
  \BibitemOpen
  \bibfield  {author} {\bibinfo {author} {\bibfnamefont {X.~M.}\ \bibnamefont
  {Chen}}, \bibinfo {author} {\bibfnamefont {D.~S.}\ \bibnamefont {Bouma}},
  \bibinfo {author} {\bibfnamefont {N.}~\bibnamefont {Kent}}, \bibinfo {author}
  {\bibfnamefont {R.}~\bibnamefont {Streubel}}, \bibinfo {author}
  {\bibfnamefont {S.}~\bibnamefont {Roy}}, \bibinfo {author} {\bibfnamefont
  {S.}~\bibnamefont {Kevan}}, \bibinfo {author} {\bibfnamefont
  {P.}~\bibnamefont {Fischer}}, \ and\ \bibinfo {author} {\bibfnamefont
  {F.}~\bibnamefont {Hellman}},\ }\href@noop {} {}\bibinfo {note} {In
  preparation}\BibitemShut {NoStop}%
\bibitem [{\citenamefont {Streubel}\ \emph {et~al.}()\citenamefont {Streubel},
  \citenamefont {Bouma}, \citenamefont {Kent}, \citenamefont {Roy},
  \citenamefont {Kevan}, \citenamefont {Fischer},\ and\ \citenamefont
  {Hellman}}]{Streubel2019}%
  \BibitemOpen
  \bibfield  {author} {\bibinfo {author} {\bibfnamefont {R.}~\bibnamefont
  {Streubel}}, \bibinfo {author} {\bibfnamefont {D.~S.}\ \bibnamefont {Bouma}},
  \bibinfo {author} {\bibfnamefont {N.}~\bibnamefont {Kent}}, \bibinfo {author}
  {\bibfnamefont {S.}~\bibnamefont {Roy}}, \bibinfo {author} {\bibfnamefont
  {S.}~\bibnamefont {Kevan}}, \bibinfo {author} {\bibfnamefont
  {P.}~\bibnamefont {Fischer}}, \ and\ \bibinfo {author} {\bibfnamefont
  {F.}~\bibnamefont {Hellman}},\ }\href@noop {} {}\bibinfo {note} {In
  preparation}\BibitemShut {NoStop}%
\bibitem [{\citenamefont {Karel}\ \emph {et~al.}(2014)\citenamefont {Karel},
  \citenamefont {Zhang}, \citenamefont {Bordel}, \citenamefont {Stone},
  \citenamefont {Chen}, \citenamefont {Jenkins}, \citenamefont {Smith},
  \citenamefont {Hu}, \citenamefont {Wu}, \citenamefont {Heald}, \citenamefont
  {Kortright},\ and\ \citenamefont {Hellman}}]{Karel2014a}%
  \BibitemOpen
  \bibfield  {author} {\bibinfo {author} {\bibfnamefont {J.}~\bibnamefont
  {Karel}}, \bibinfo {author} {\bibfnamefont {Y.~N.}\ \bibnamefont {Zhang}},
  \bibinfo {author} {\bibfnamefont {C.}~\bibnamefont {Bordel}}, \bibinfo
  {author} {\bibfnamefont {K.~H.}\ \bibnamefont {Stone}}, \bibinfo {author}
  {\bibfnamefont {T.~Y.}\ \bibnamefont {Chen}}, \bibinfo {author}
  {\bibfnamefont {C.~A.}\ \bibnamefont {Jenkins}}, \bibinfo {author}
  {\bibfnamefont {D.~J.}\ \bibnamefont {Smith}}, \bibinfo {author}
  {\bibfnamefont {J.}~\bibnamefont {Hu}}, \bibinfo {author} {\bibfnamefont
  {R.~Q.}\ \bibnamefont {Wu}}, \bibinfo {author} {\bibfnamefont {S.~M.}\
  \bibnamefont {Heald}}, \bibinfo {author} {\bibfnamefont {J.~B.}\ \bibnamefont
  {Kortright}}, \ and\ \bibinfo {author} {\bibfnamefont {F.}~\bibnamefont
  {Hellman}},\ }\href {\doibase 10.1088/2053-1591/1/2/026102} {\bibfield
  {journal} {\bibinfo  {journal} {Mater. Res. Express}\ }\textbf {\bibinfo
  {volume} {1}},\ \bibinfo {pages} {026102} (\bibinfo {year}
  {2014})}\BibitemShut {NoStop}%
\bibitem [{\citenamefont {Spaldin}(2010)}]{Spaldin2010}%
  \BibitemOpen
  \bibfield  {author} {\bibinfo {author} {\bibfnamefont {N.~A.}\ \bibnamefont
  {Spaldin}},\ }\href@noop {} {\emph {\bibinfo {title} {Magnetic Materials}}}\
  (\bibinfo  {publisher} {Cambridge University Press},\ \bibinfo {year}
  {2010})\BibitemShut {NoStop}%
\bibitem [{\citenamefont {Stoner}(1938)}]{Stoner1938}%
  \BibitemOpen
  \bibfield  {author} {\bibinfo {author} {\bibfnamefont {E.~C.}\ \bibnamefont
  {Stoner}},\ }\href {\doibase 10.1098/rspa.1938.0066} {\bibfield  {journal}
  {\bibinfo  {journal} {Proc. R. Soc. London. Ser. A. Math. Phys. Sci.}\
  }\textbf {\bibinfo {volume} {165}},\ \bibinfo {pages} {372} (\bibinfo {year}
  {1938})}\BibitemShut {NoStop}%
\bibitem [{\citenamefont {Karel}\ \emph {et~al.}(2018)\citenamefont {Karel},
  \citenamefont {Bouma}, \citenamefont {Martinez}, \citenamefont {Zhang},
  \citenamefont {Gifford}, \citenamefont {Zhang}, \citenamefont {Zhao},
  \citenamefont {Kim}, \citenamefont {Li}, \citenamefont {Huang}, \citenamefont
  {Wu}, \citenamefont {Chen},\ and\ \citenamefont {Hellman}}]{Karel2018}%
  \BibitemOpen
  \bibfield  {author} {\bibinfo {author} {\bibfnamefont {J.}~\bibnamefont
  {Karel}}, \bibinfo {author} {\bibfnamefont {D.~S.}\ \bibnamefont {Bouma}},
  \bibinfo {author} {\bibfnamefont {J.}~\bibnamefont {Martinez}}, \bibinfo
  {author} {\bibfnamefont {Y.~N.}\ \bibnamefont {Zhang}}, \bibinfo {author}
  {\bibfnamefont {J.~A.}\ \bibnamefont {Gifford}}, \bibinfo {author}
  {\bibfnamefont {J.}~\bibnamefont {Zhang}}, \bibinfo {author} {\bibfnamefont
  {G.~J.}\ \bibnamefont {Zhao}}, \bibinfo {author} {\bibfnamefont {D.~R.}\
  \bibnamefont {Kim}}, \bibinfo {author} {\bibfnamefont {B.~C.}\ \bibnamefont
  {Li}}, \bibinfo {author} {\bibfnamefont {Z.~Y.}\ \bibnamefont {Huang}},
  \bibinfo {author} {\bibfnamefont {R.~Q.}\ \bibnamefont {Wu}}, \bibinfo
  {author} {\bibfnamefont {T.~Y.}\ \bibnamefont {Chen}}, \ and\ \bibinfo
  {author} {\bibfnamefont {F.}~\bibnamefont {Hellman}},\ }\href {\doibase
  10.1103/PhysRevMaterials.2.064411} {\bibfield  {journal} {\bibinfo  {journal}
  {Phys. Rev. Mater.}\ }\textbf {\bibinfo {volume} {2}},\ \bibinfo {pages}
  {064411} (\bibinfo {year} {2018})}\BibitemShut {NoStop}%
\end{thebibliography}

%

\end{document}